\documentclass[fleqn,usenatbib]{mnras}

\usepackage{newtxtext,newtxmath}

\usepackage[T1]{fontenc}
\usepackage{ae,aecompl}

\usepackage{graphicx}	
\usepackage{amsmath}	
\usepackage{hyperref}
\usepackage{changes}
\usepackage{xcolor}
\definechangesauthor[name={Bibiana Prinoth}, color=cyan]{BP}
\definechangesauthor[name={Elspeth Lee}, color=red]{EL}
\usepackage{xspace}
\newcommand{\kmpers}{\,km\,s$^{-1}$\xspace}
\newcommand{\kpvsys}{$K_p-V_{\rm sys}$\xspace}

\defcitealias{Hammond2022}{Hammond et al.}

\def\apj{ApJ}

\def\araa{ARA\&A}
\def\aap{A\&A}
\def\apjl{ApJL}
\def\apjs{ApJS}
\def\mnras{MNRAS}
\def\aj{AJ}
\def\nat{Nature}

\def\icarus{Icarus}

\hypersetup{
  draft=false,
	pdftoolbar=false,
	pdfmenubar=true,
	pdfstartview={FitP},
	breaklinks=false,
	frenchlinks=false,
  colorlinks=true,
  linkcolor=red,
  citecolor=blue,
  filecolor=cyan,
  urlcolor=magenta,
  pdftitle={},
	pdfsubject={},
	pdfauthor={Elspeth K. H. Lee}
  }

\DeclareRobustCommand{\VAN}[3]{#2} 

\title[3D Hi-res UHJ]{\centering{The Mantis Network \texttt{II}: \\
Examining the 3D high-resolution observable properties of the UHJs WASP-121b and WASP-189b through GCM modelling}}

\author[Lee et al.]{Elspeth K. H. Lee$^{1}$, Bibiana Prinoth$^{2}$, Daniel Kitzmann$^{1}$, Shang-Min Tsai$^{3}$,  \newauthor
Jens Hoeijmakers$^{2}$ , Nicholas W. Borsato$^{2}$ and Kevin Heng$^{1,4,5}$ \\
$^{1}$Center for Space and Habitability, University of Bern, Gesellschaftsstrasse 6, CH-3012 Bern, Switzerland \\
$^{2}$Lund Observatory, Department of Astronomy and Theoretical Physics, Lund University, Box 43, 221 00 Lund, Sweden \\
$^{3}$Atmospheric, Oceanic \& Planetary Physics, Department of Physics, University of Oxford, Oxford OX1 3PU, UK \\
$^{4}$University of Warwick, Department of Physics, Astronomy \& Astrophysics Group, Coventry CV4 7AL, UK \\
$^{5}$Ludwig Maximilian University, University Observatory Munich, Scheinerstrasse 1, Munich D-81679, Germany.
}

\date{Accepted 2022 August 8. Received 2022 June 20; in original form 2022 August 1}

\pubyear{2022}

\begin{document}
\label{firstpage}
\pagerange{\pageref{firstpage}--\pageref{lastpage}}
\maketitle

\begin{abstract}
The atmospheres of ultra hot Jupiters (UHJs) are prime targets for the detection of molecules and atoms at both low and high spectral resolution.
We study the atmospheres of the UHJs WASP-121b and WASP-189b by performing 3D general circulation models (GCMs) of these planets using high temperature correlated-k opacity schemes with ultra-violet (UV) absorbing species included.
The GCM results are then post-processed at low and high spectral resolutions and compared to available data.
The high resolution results are cross-correlated with molecular and atomic templates to produce mock molecular detections.
Our GCM models produce similar temperature-pressure (T-p) structure trends to previous 1D radiative-convective equilibrium models of UHJs.
Furthermore, the inclusion of UV opacities greatly shapes the thermal and dynamical properties of the high-altitude, low-pressure regions of the UHJ atmospheres, with sharp T-p inversions due to the absorption of UV light.
This suggests that optical wavelength, high-resolution observations probe a dynamically distinct upper atmospheric region, rather than the deeper jet forming layers.
\end{abstract}




\section{Introduction}
\label{sec:intro}

Observational programs of ultra hot Jupiters (UHJs) continue to produce numerous detections of molecules, determination of temperature structures and measurements of wind speeds in their atmospheres.
Due to their high temperatures and large scale heights, UHJs have favourable properties for strong signals of their atmospheres to be observed in transmission and emission.
Two UHJs that have been extensively observed are WASP-121b and WASP-189b, which are prime targets for low and high resolution spectral observations.

WASP-121b \citep{Delrez2016} is an UHJ that orbits a F type star with an equilibrium temperature of T$_{\rm eq}$ $\approx$ 2360 K.
WASP-121b is scheduled for full orbit phase curve observations for the cycle 1 JWST GTO program with NIRISS \citep{Lafreniere2017_JWST} and a GO program with the NIRSpec instrument \citep{Evans2021_JWST}.
These observations, combined with the available optical and near-IR data from HST and ground based telescopes \citep{Evans2020, Wilson2021} will provide some of the most detailed and wavelength expansive atmospheric characterisation of a UHJ to date.
Recently, \citet{Bourrier2020} and \citet{Daylan2021} produced TESS photometric phase curves of WASP-121b and \citet{Mikal-Evans2022} observed full phase orbits using HST WFC3.
In \citet{Parmentier2018} and \citet{Mikal-Evans2022}, WASP-121b was modelled using the SPARC/MITgcm 3D GCM model, finding that H$_{2}$O dissociation and H$^{-}$ opacities were important considerations when modelling UHJ atmospheres to fit their observed emission spectra.

Numerous ground based high-resolution observations have targeted WASP-121b \citep[e.g.][]{Merritt2020, Gibson2020, Hoeijmakers2020, Merritt2021}, resulting in various detections (and non-detections) of molecules and atoms in its atmosphere.

WASP-189b \citep{Anderson2018} is an UHJ that orbits an A type star with an equilibrium temperature of T$_{\rm eq}$ $\approx$ 2620 K.
\citet{Prinoth2022} cross-correlated HARPS and HARPS-N transmission data of WASP-189b and were able to detect TiO, Fe, Fe+, Ti, Ti+, Cr, Mg, V and Mn.
However, the K$_{\rm p}$-V$_{\rm sys}$ signal from each species was different, suggesting chemical inhomogeneity in the atmosphere.
\citet{Lendl2020} and \citet{Deline2022} present CHEOPS photometric data of WASP-189b including an optical phase curve.

In this study, we investigate the 3D dynamical and thermal properties of the atmospheres of WASP-121b and WASP-189b and how this manifests in mock high-resolution spectral observations.
We perform GCM models of both planets using Exo-FMS \citep{Lee2021} coupled to a correlated-k radiative-transfer scheme with updated species that absorb at UV wavelengths.
We then post-process the results of the GCM modelling at low and high resolution using the 3D gCMCRT Monte Carlo radiative-transfer model \citep{Lee2022b}.
The low-resolution results are then compared to the available transmission, emission and phase curve observational data of each planet.
The high-resolution results are used in mock cross-correlation calculations to produce synthetic molecular and atomic detections from our GCM output.

In Section \ref{sec:GCM_modelling} we provide details on the GCM modelling components of each planet.
Section \ref{sec:opacity} details the correlated-k opacity scheme with UV absorbing species used in the GCM modelling.
Section \ref{sec:GCM_pp} details the low and high resolution post-processing performed on the GCM output to required to produce the synthetic observations.
Section \ref{sec:GCM_results} shows the results of the WASP-121b and WASP-189b GCM model.
Section \ref{sec:low-res} presents the low-resolution spectral post-processing results and the comparison to the available transmission, emission and phase curve data on both objects
Section \ref{sec:high-res} presents the high-resolution cross-correlation results of both simulated planets.
Section \ref{sec:discussion} contains the discussion about our GCM results and post-processing efforts.
Section \ref{sec:conclusions} contains the summary and conclusions of our study.

\section{GCM modelling}
\label{sec:GCM_modelling}

\begin{table*}
\centering
\caption{Adopted GCM simulation parameters for WASP-121b and WASP-189b.}
\begin{tabular}{c c c c l}  \hline \hline
 Symbol & WASP-121b & WASP-189b & Unit & Description \\ \hline
 T$_{\rm int}$ & 583 & 497 & K & Internal temperature \\
 P$_{\rm 0}$ & 1000 & 1000 &  bar & Reference surface pressure \\
 c$_{\rm P}$ & 13000 & 13000 &  J K$^{-1}$ kg$^{-1}$ & Specific heat capacity \\
 R & 3556.8 & 3556.8 &  J K$^{-1}$ kg$^{-1}$  & Ideal gas constant \\
 $\kappa$ & 0.2736 & 0.2736 & -  & Adiabatic coefficient \\
 g$_{\rm p}$ & 8.44 & 18.82 & m s$^{-2}$ & Acceleration from gravity \\
 R$_{\rm p}$ & 1.33 $\times$ 10$^{8}$ & 1.16 $\times$ 10$^{8}$ & m & Radius of planet \\
 $\Omega_{\rm p}$ & 5.70 $\times$ 10$^{-5}$ &  2.67 $\times$ 10$^{-5}$& rad s$^{-1}$ & Rotation rate of planet \\
 $\Delta$ t$_{\rm hydro}$ & 20 & 20 & s & Hydrodynamic time-step \\
 $\Delta$ t$_{\rm rad}$  & 160 & 160 & s & Radiative time-step \\
 N$_{\rm v}$ & 54 & 54 & - & Vertical resolution \\
 d$_{\rm 2}$ & 0.02 & 0.02 & - & div. dampening coefficient \\
\hline
\end{tabular}
\label{tab:GCM_parameters}
\end{table*}

For our 3D modelling of WASP-121b and WASP-189b we use the Exo-FMS GCM model in the gas giant atmosphere set-up \citep[e.g.][]{Lee2021}.
The parameters used for both planets are detailed in Table \ref{tab:GCM_parameters}.
Both simulations were performed with a cubed-sphere C48 ($\approx$ 192 x 96 in longitude and latitude elements) resolution grid.
We use 54 vertical layers between 10$^{-6}$-1000 bar.

To help stabilise the simulations in the deep atmosphere, we include a linear Rayleigh `basal' drag \citep{Liu2013, Komacek2016, Tan2019, Carone2020} which applies between 500 to 1000 bar, with a base drag timescale of 1 Earth day.
This can be physically motivated as a deep magnetic drag acting on the atmosphere, however, the exact drag timescale is chosen for minimal stabilisation reasons, rather than based on physical justifications \citep[e.g. see][]{Beltz2022}.
Numerical stabilisation in Exo-FMS is provided by a second order divergence dampening with coefficient d$_{2}$ = 0.02.
We do not include upper atmospheric drag as a boundary condition, for example, reducing the zonal velocities to zero, rather Exo-FMS increases the strength of the divergence dampening at the upper boundary.
This and other numerical stabilisation schemes used in other contemporary GCM models are discussed in depth in \citetalias{Hammond2022} (\citeyear{Hammond2022}).

For the two stream column-wise radiative-transfer (RT) within the GCM, we use the short characteristics method \citep{Olson1987} for the longwave internal radiation, and for the shortwave incident stellar flux we use the `adding method' \citep[e.g.][]{Mendonca2015}, which is a fast method that includes the effects of scattering.
We increase the accuracy of the longwave RT scheme by using Bezier interpolation to vertically interpolate the temperature from the layers (centre of the cells) to the levels (edges of the cells) as described in \citet{Lee2022}.
The RT is calculated every 8 hydrodynamical timesteps, giving a radiative-timestep of 160 seconds compared to the hydrodynamical timestep of 20 seconds.
This is required to balance accurately capturing the small radiative-timescales at the top of the UHJ atmospheres and the computational efficiency of the model integration.

Our approach is a fast and efficient method for performing the two-stream calculations inside the GCM\footnote{The radiative-transfer models used in this study are available publicly on GitHub: \url{https://github.com/ELeeAstro}}.
However, we neglect the scattering for longwave radiation (thermal radiation inside the GCM columns).
The effect of scattering in the longwave is likely to be unimportant unless clouds or hazes are present in the atmospheric column.
This likely the case for some UHJ planets \citep[e.g.][]{Helling2021,Komacek2022}, with clouds mostly confined to the nightside hemisphere.
However, due to the significant added computational expense of modelling clouds in hot Jupiter GCMs \citep[e.g.][]{Lee2016, Lines2018,Roman2021, Christie2021}, we do not consider the formation of clouds or hazes inside the GCM.

Following the scheme in \citet{Lee2021}, we run each simulation for 2000 days using the `picket fence' RT scheme \citep{Parmentier2015, Lee2021} for an extended spin-up period.
The RT scheme is then switched to the full correlated-k treatment for another 1000 days of simulation.
We then simulate for a further 50 days, taking this average as the `final' result.

To aid the radiative-convective equilibrium properties of the deep atmosphere, we follow the scheme outlined in \citet{Lee2021} where the analytical T-p profile from \citet{Parmentier2015} with adiabatic correction at the sub-stellar point is used as the initial conditions.
For the internal heat, we use the relation in \citet{Thorngren2019} as the internal temperature, T$_{\rm int}$ (= 583 K for WASP-121b and = 497 K for WASP-189b) for each planet, both for the initial conditions and as the internal heat flux in the RT scheme during GCM runtime.
This scheme allows the deep adiabatic region to cool quickly to the `correct' gradient, following the advice in \citet{Sainsbury-Martinez2019}.

For the stellar incident flux in the correlated-k radiative-transfer module, we use PHOENIX stellar atmosphere models \citep{Allard2012} interpolated to the properties of WASP-121 (T$_{\rm eff}$ = 6460.0 K, [Fe/H] = 0.13, log g = 3.9) and WASP-189 (\citet{Deline2022}: T$_{\rm eff}$ = 8000 K, [Fe/H] = 0.29, log g = 3.9) using pysnphot \citep{pysynphot2013}.

We also include the effects of non-constant heat capacity on the RT heating rates.
We adopt a module\footnote{This scheme can be found on the lead author's GitHub: \url{https://github.com/ELeeAstro}} that interpolates the H$_{2}$, H and He heat capacities from the JANAF tables \citep{Chase1986} as a function of temperature.
The heat capacity of the cell is then given by the mass ratio weighted abundances of the three species.

\section{Opacity sources}
\label{sec:opacity}

\begin{table}
\centering
\caption{Opacity sources and references used in the Exo-FMS GCM correlated-k scheme and gCMCRT post-processing. `ff' denotes free-free opacity.}
\begin{tabular}{c l}  \hline \hline
Opacity Source & Reference  \\ \hline
Atomic &   \\ \hline
Na  &  \citet{Kurucz1995}   \\
K   &  \citet{Kurucz1995}  \\
Fe   &  \citet{Kurucz1995}  \\
Fe$^{+}$   &  \citet{Kurucz1995}  \\ \hline
Molecular &  \\ \hline
H$_{2}$O   &  \citet{Polyansky2018}   \\
OH   &  \citet{Hargreaves2019}   \\
CH$_{4}$   &  \citet{Hargreaves2020}  \\
C$_{2}$H$_{2}$   &  \citet{Chubb2020}   \\
C$_{2}$ & \citet{Yurchenko2018} \\
CH   &   \citet{Masserson20214} \\
CN   &   \citet{Syme2021} \\
CO   &  \citet{Li2015}   \\
CO$_{2}$    &  \citet{Yurchenko2020}   \\
NH$_{3}$   &  \citet{Coles2019}   \\
PH$_{3}$ & \citet{Sousa_Silva2015} \\
H$_{2}$S   & \citet{Azzam2016} \\
HCl   & \citet{Gordon2017} \\
HCN   & \citet{Harris2006, Barber2014} \\
SH & \citet{Gorman2019} \\
HF & \citet{Li2013,Coxon2015} \\
N$_{2}$ & \citet{Western2018} \\
H$_{2}$ & \citet{Roueff2019} \\
SiO & \citet{Yurchenko2021} \\
TiO & \citet{McKemmish2019} \\
VO & \citet{McKemmish2016} \\
MgH & \citet{GharibNezhad2013} \\
CaH & \citet{Bernath2020} \\
TiH & \citet{Bernath2020} \\
CrH & \citet{Bernath2020} \\
FeH & \citet{Bernath2020} \\ \hline
Continuum  & \\ \hline
H$_{2}$-H$_{2}$  &  \citet{Karman2019} \\
H$_{2}$-He   & \citet{Karman2019} \\
H$_{2}$-H   & \citet{Karman2019} \\
H-He   & \citet{Karman2019} \\
H$^{-}$ & \citet{John1988} \\
H$_{2}^{-}$ (ff)  & \citet{Bell1980} \\
He$^{-}$ (ff)  & \citet{Bell1982} \\ \hline
Rayleigh scattering & \\ \hline
H$_{2}$ & \citet{Dalgarno1962} \\
He  & \citet{Thalman2014} \\
H  & \citet{Kurucz1970} \\
e$^{-}$ & Thomson scattering\\ \hline
\end{tabular}
\label{tab:line-lists}
\end{table}

Table \ref{tab:line-lists} provides the list of opacity sources, including 30 atomic/molecular species.
Specifically, we include the important UV-optical absorber species TiO, VO, SiO, Fe and Fe$^{+}$, responsible for generating the upper atmospheric temperature inversion in high temperature atmospheres \citep[e.g.][]{Lothringer2020b}.
For completeness, we include carbon molecules known to be important for the cooling of lower temperature stars, namely CN, CH and C$_{2}$, which are usually more abundant at higher C/O ratios.
Our pre-mixed k-table is valid between a temperature of 200-6100 K and pressures between 10$^{-8}$-1000 bar, suitable for the range of temperatures and pressures for UHJ atmospheric modelling.
We use the same 32 wavelength band scheme as in \citet{Amundsen2014}, which has a minimum wavelength edge of 0.2 $\mu$m, capturing most of the near-UV absorption in the atmosphere.
During runtime we combine the atomic and molecular line opacities with the continuum and Rayleigh scattering opacity sources.

To construct `pre-mixed' k-tables, the volume mixing ratio of each species is required to be known \citep{Amundsen2017}.
For this, we use the GGchem chemical equilibrium (CE) code \citep{Woitke2018}, including condensation and thermal ionisation of species.
In addition, during runtime we interpolate to a 2D grid in temperature and pressure produced using GGchem to find the mean molecular weight and volume mixing ratios (VMRs) for the collision induced absorption (CIA) and Rayleigh scattering species in the opacity scheme.
Solar metallicity is assumed for these calculations using the Solar elemental ratios presented in \citet{Asplund2009}.

\section{GCM post-processing}
\label{sec:GCM_pp}

To post-process the results from the GCM we use the 3D gCMCRT Monte Carlo radiative-transfer code \citep{Lee2022b}.
For all models we use the same sources for the opacities as in Table \ref{tab:line-lists}.
The gCMCRT framework has been used in previous studies to investigate the properties of 3D atmospheric models at high spectral resolution \citep[e.g.][]{Wardenier2021,Wardenier2022,vanSluijs2022}.

At low-resolution, we perform all post-processing at a resolution of R\,$\approx$\,100, suitable for producing synthetic JWST instrument predictions from the model output.
We use the correlated-k method, with 503 bins between a wavelength range of 0.2-30\,$\mu$m.
Individual species k-tables are combined using the random overlap method with resort and rebin \citep{Amundsen2017}.

At high wavelength resolutions, we target the ESPRESSO instrument \citep{Pepe2021} wavelength range (0.378-0.789 $\mu$m), and perform the post-processing at R = 140,000, for a total of $\approx$ 110,000 individual wavelength points.
These simulations cover half an orbit of WASP-121b across the secondary eclipse, to match observations made by Hoeijmakers et al. (in prep) that show several species in the emission spectrum.
The phase coverage of the observations was $\phi$ = 0.25 - 0.75 (secondary eclipse), with the resolution of observations taken at R = 140,000.
We therefore mimic this observational run by performing the post-processing of GCM for 10 phases between these values.
We include line opacities from the species H$_{2}$O, OH, CO, SiO, Fe, Fe$^{+}$, TiO, Ti, Ti$^{+}$, VO, V, V$^{+}$ as well as continuum opacity from H$^{-}$ and Rayleigh scattering from H$_{2}$, He and H.
For the stellar fluxes, we use PHOENIX models interpolated to the stellar properties of WASP-121 using pysnphot \citep{pysynphot2013}, the same as in Sect. \ref{sec:GCM_modelling}.

\citet{Prinoth2022} detected Fe, Fe$^+$, Ti, Ti$^+$, Cr, Mg, V, Mn and TiO and tentatively detected Cr$^+$, Sc$^+$, Na, Ni and Ca in the transmission spectrum of WASP-189b by combining five transit time series obtained with the HARPS \citep{Mayor2003} and HARPS-N \citep{cosentino2012harps} spectrographs. Two of the five time series do not cover the entire transit, which lies between the phases $\phi = 0.967 - 0.033$, because the observations were only started after ingress.

We follow the observation strategy in \citet{Prinoth2022} for our synthetic spectra modelling, producing transit spectra across the transmission phases at a resolution of R\,=\,120,000 between 0.378 - 0.691\,$\mu$m for a total of $\approx$ 72,000 wavelength points.
The same as our WASP-121b models, we include opacities from H$_{2}$O, OH, CO, SiO, Fe, Fe$^{+}$, TiO, Ti, Ti$^{+}$, VO, V, V$^{+}$, continuum H$^{-}$ and H$_{2}$, He and H Rayleigh scattering.

To include the effect of limb darkening in the HARPS band we use a quadratic darkening law, with coefficients given by the CHEOPS observations from \citet{Deline2022} with c$_{1}$ = 0.414 and c$_{2}$ = 0.155.
We use the spherical geometric limb-darkening scheme in \citet{Lee2022b} to calculate the effect of limb darkening on the transmission spectra.
We take the inclination of the planet during transit to be constant at 84.58$^{\circ}$ \citep{Deline2022}.

\section{GCM results}
\label{sec:GCM_results}

In this section, we present the results of our Exo-FMS WASP-121b and WASP-189b GCM simulations.

\subsection{WASP-121b GCM}

\begin{figure*} 
   \centering
   \includegraphics[width=0.49\textwidth]{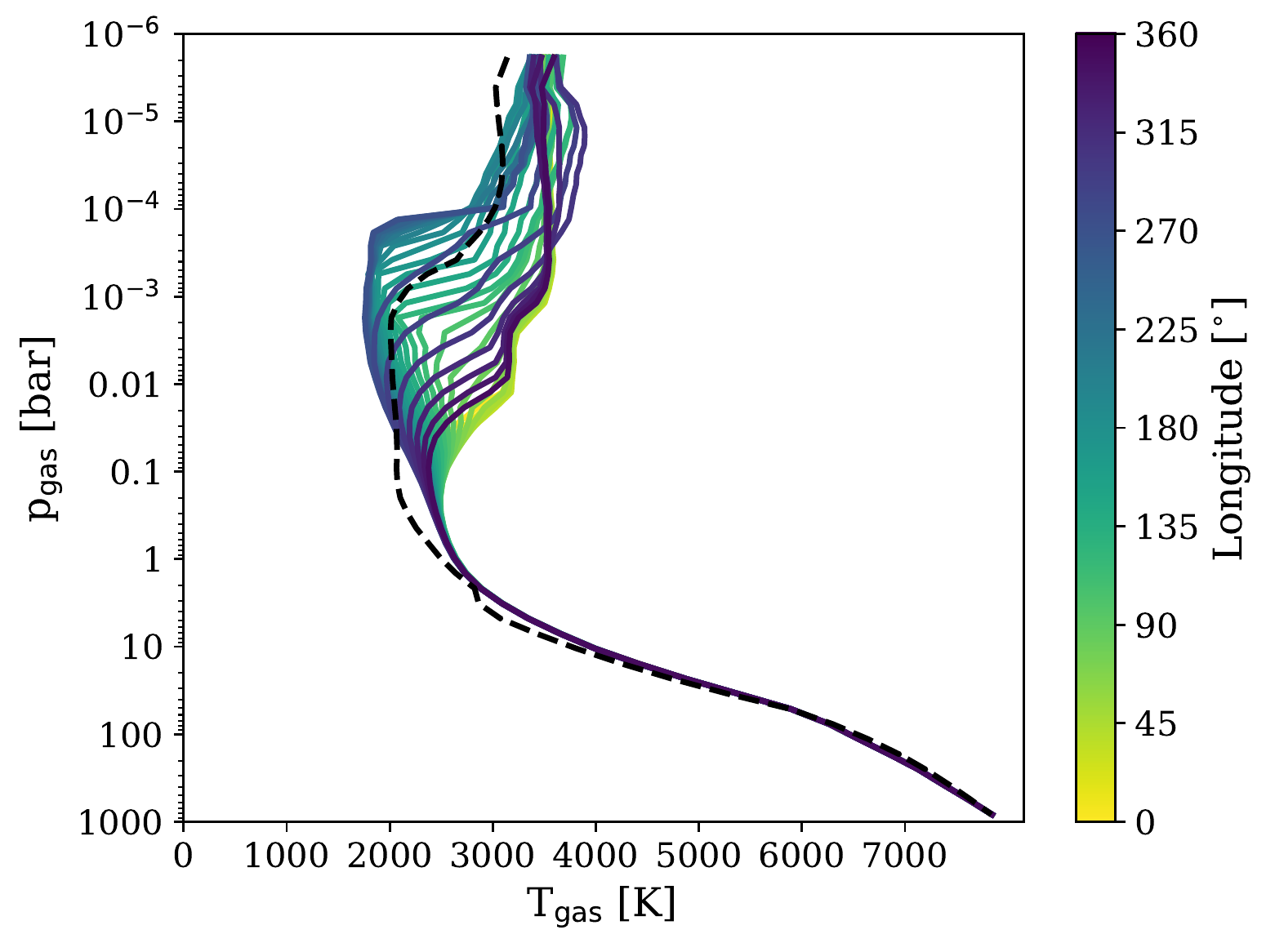}
   \includegraphics[width=0.49\textwidth]{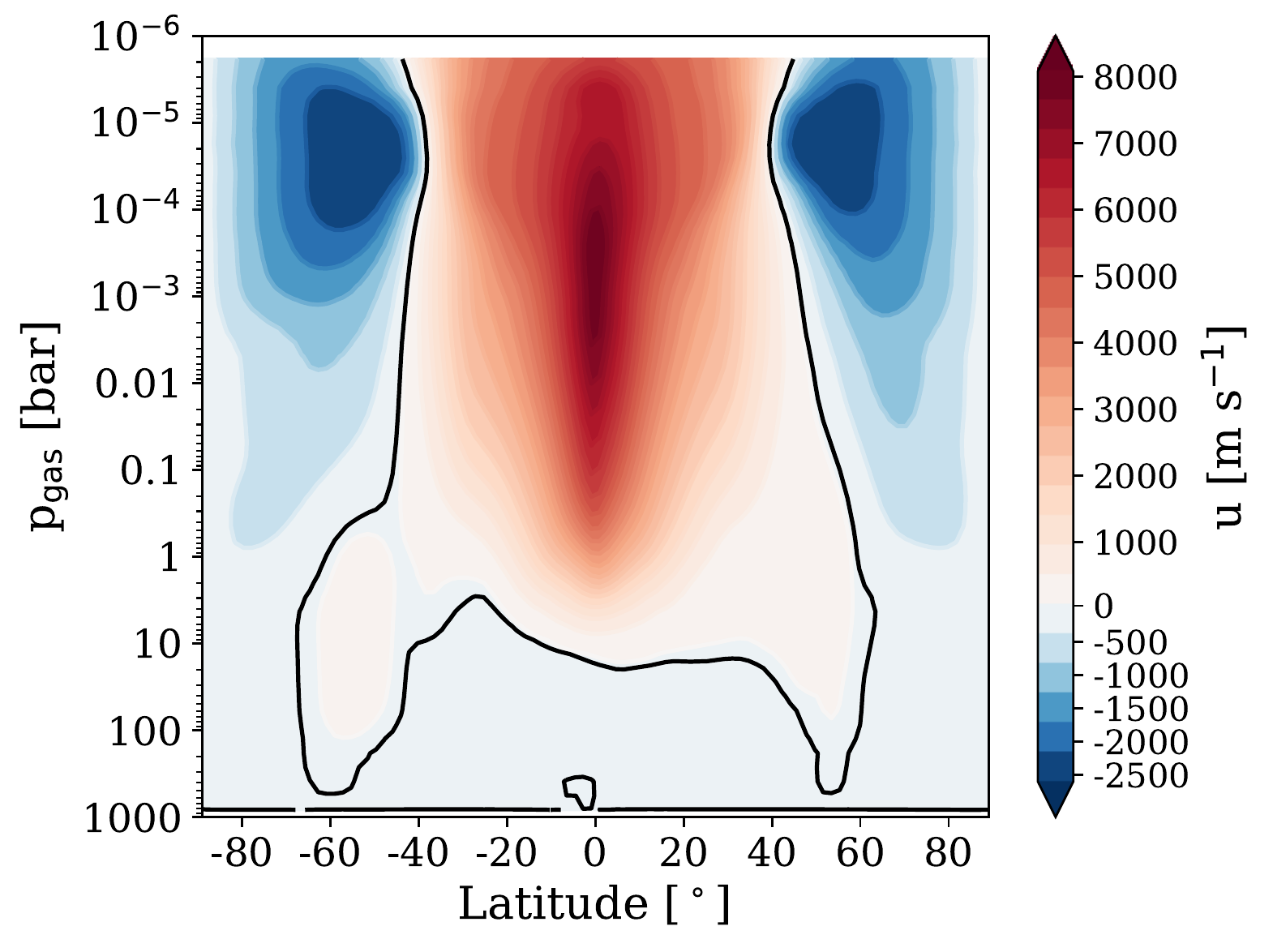}
   \includegraphics[width=0.49\textwidth]{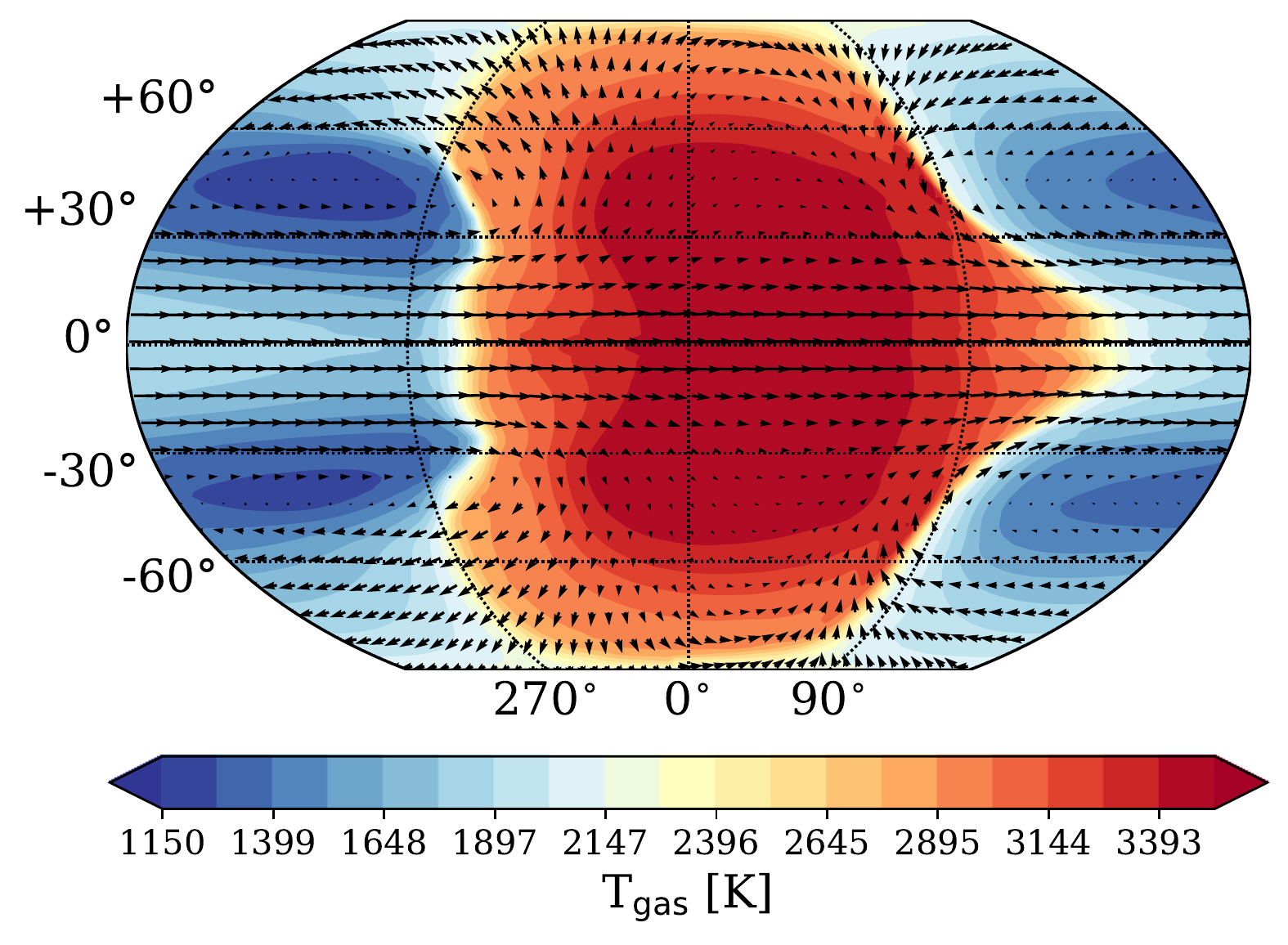}
   \includegraphics[width=0.49\textwidth]{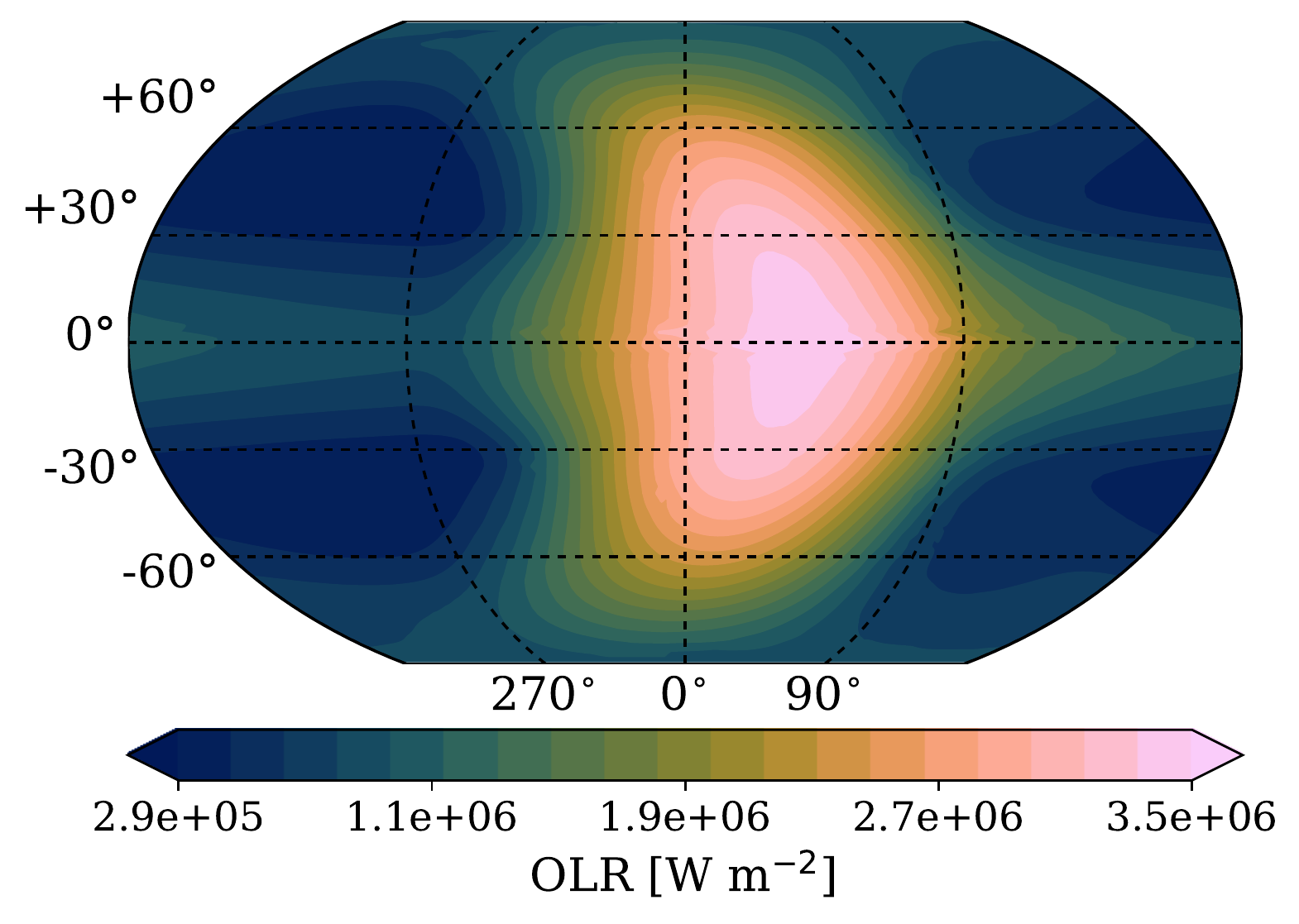}
   \caption{WASP-121b GCM results from Exo-FMS.
   Top left: Vertical T-p profiles at the equatorial regions (solid coloured lines) and polar column (dashed).
   Top right: Zonal mean zonal velocity.
   Bottom left: Temperature map at 1 mbar.
   Bottom right: Outgoing longwave radiation (OLR) map.}
   \label{fig:WASP121_GCM}
\end{figure*}

In Figure \ref{fig:WASP121_GCM} we show the results of our WASP-121b GCM model.
Our produced T-p structure shows a typical ultra hot Jupiter temperature structure, with a strong upper atmospheric temperature inversion, in line with previous studies \citep[e.g.][]{Parmentier2018}.
The slight kink in the T-p profile around 6000 K is probably due to the temperature passing the maximum value of the k-tables at 6100 K.
Beyond this temperature the opacities are forced to be at the 6100 K values, leading to a slight deficit in the opacity distribution due to not capturing the extra thermal broadening of the lines.
However, we calculate the global average radiative-convective boundary pressure level in the GCM model to be $\approx$3.7 bar, showing these deep layers should be convectively driven rather than radiative.
The steep temperature inversion at low pressures ($\approx$ 10$^{-4}$ bar) is similar to 1D radiative-convective equilibrium models performed in \citet{Lothringer2020b}, showing the effect of adding high altitude UV absorbing species on the T-p structure.
This forcing from the UV radiation at very low pressure homogenizes the temperature across the hemispheres through driving strong low-pressure winds, suggesting highly efficient energy transport occurring across all hemispheres of the planet at these pressures.
This suggests that this region is somewhat decoupled dynamically from the lower atmosphere, which shows a stronger day-night contrast variation with pressure.

The zonal mean zonal velocity plots suggest very strong, several km s$^{-1}$ winds at the equatorial regions of the planet.
Due to the additional high altitude forcing from the UV absorbers, this strong jet extends to very low pressure, typical regions where high-resolution spectroscopy at optical wavelengths would probe and strong lines from UV-optical absorbing species would be manifest.
Our 3D temperature map and outgoing longwave radiation (OLR) plot in Fig. \ref{fig:WASP121_GCM} show an eastward offset in the maximum outgoing flux, a common feature of hot Jupiter models \citep[e.g.][]{Heng2011}.

\subsubsection{Comparison to previous GCMs}

\begin{figure} 
   \centering
   \includegraphics[width=0.49\textwidth]{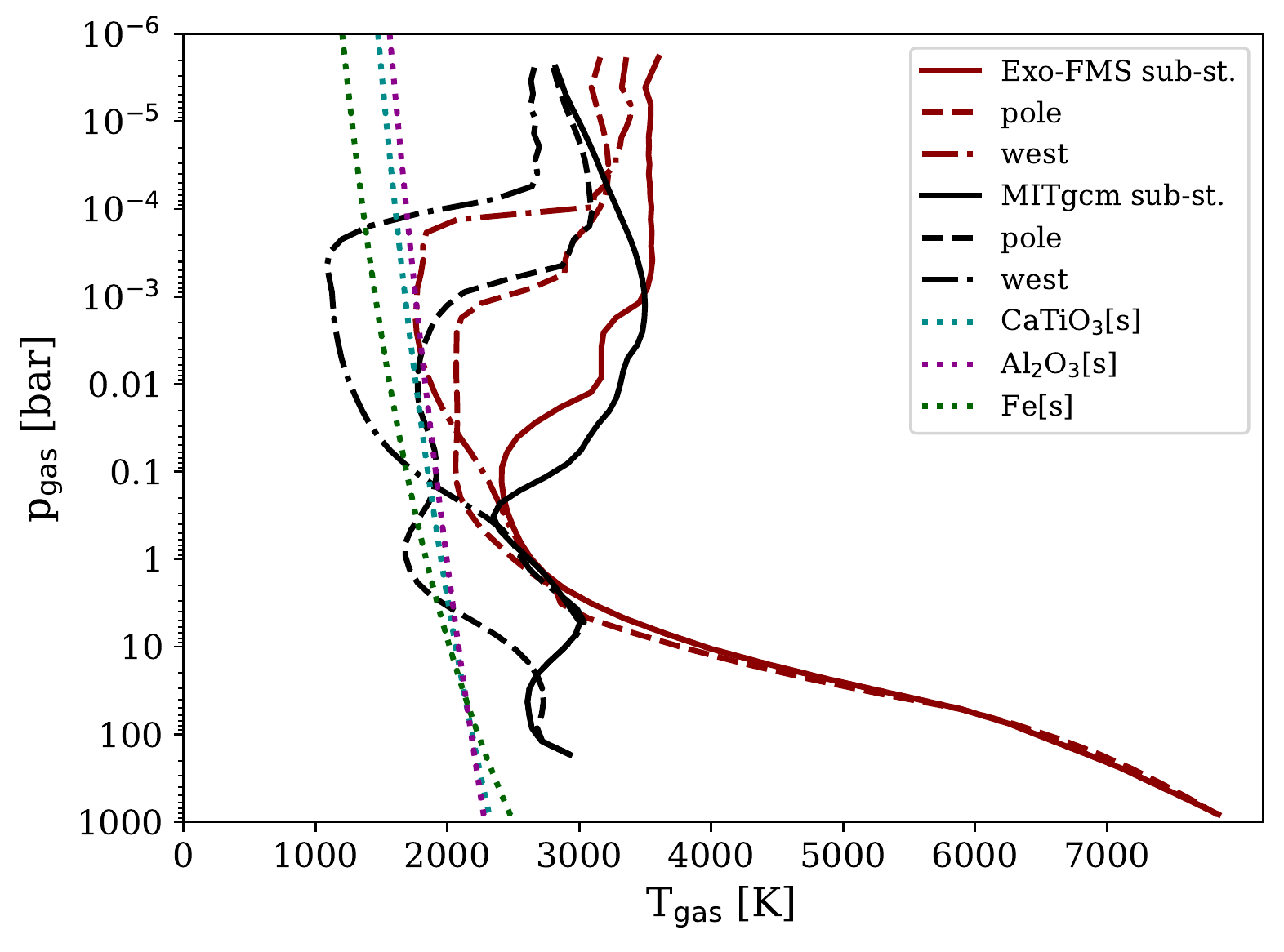}
   \caption{WASP-121b GCM T-p profiles of the sub-stellar point (solid lines), pole (dashed lines) and western limb (dash-dot) from the Exo-FMS (red) and SPARC/MITgcm (black).
   Overplotted are the condensation curves for solid CaTiO$_{3}$, Al$_{2}$O$_{3}$ and Fe.}
   \label{fig:WASP121_GCM_comp}
\end{figure}

In \citet{Parmentier2018} a GCM using SPARC/MITgcm was performed for WASP-121b.
We briefly compare the WASP-121b results between the models.
Figure \ref{fig:WASP121_GCM_comp} shows the T-p profiles of the sub-stellar, polar and west terminator profiles of both models.
The major differences are seen in the deep atmosphere, where the Exo-FMS deep region follows a more high internal temperature adiabatic profile compared to the more isothermal SPARC/MITgcm, probably due to the different assumptions of the initial T-p conditions and internal temperature used between the models.
Differences due to the inclusion of UV opacities in the Exo-FMS RT scheme are clear at low pressures $<$ 10$^{-4}$ bar, with Exo-FMS inverting towards more isothermal temperatures following the behaviour found in the \citet{Lothringer2020b} 1D RCE models, while the SPARC/MITgcm inversion drops off in temperature from 10$^{-3}$.
This T-p profile behaviour is similar to that found in \citet{Lee2021} for an HD 209458b simulation, which also did not include UV absorbers.

\subsection{WASP-189b GCM}

Our WASP-189b GCM results are similar to the WASP-121b results, but with a steeper low pressure temperature inversion.
The dayside temperature structures are again reminiscent of the 1D modelling of \citet{Lothringer2020b}, with more isothermal temperatures occurring at very low pressure due to highly efficient energy transport.
At deeper pressures we also see a stronger day/night temperature contrast, probably driven by the overall larger dayside temperatures resulting in a lower radiative timescale.
This is reflected in the OLR plot, with less of an eastward shift seen in the WASP-189b model compared to WASP-121b.
Again, we see the kink in the T-p profile near 6100 K, where the opacity tables reach their maximum temperatures.
We calculate the global average radiative-convective boundary pressure level in the GCM model to be $\approx$5.1 bar.

\begin{figure*} 
   \centering
   \includegraphics[width=0.49\textwidth]{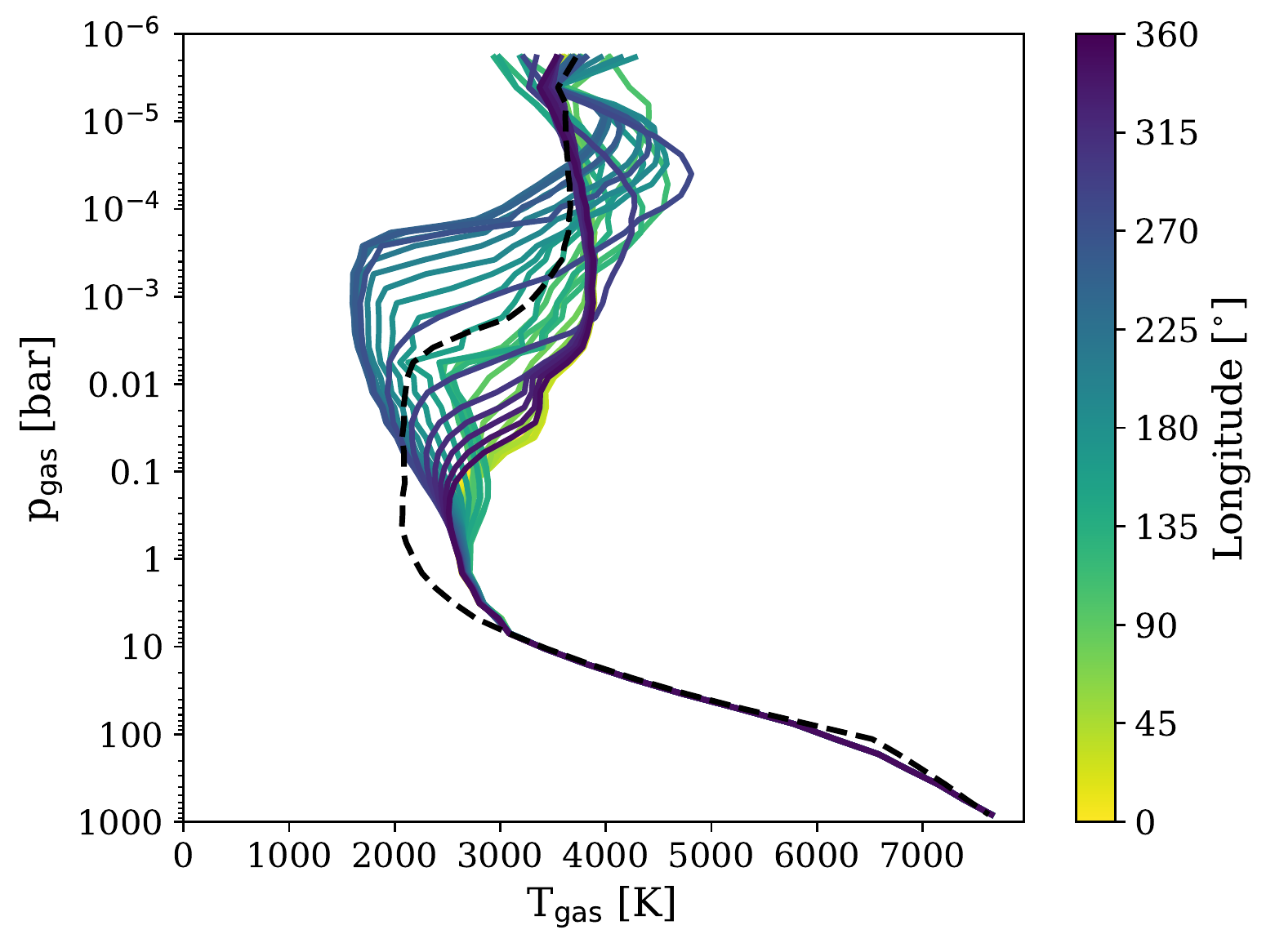}
   \includegraphics[width=0.49\textwidth]{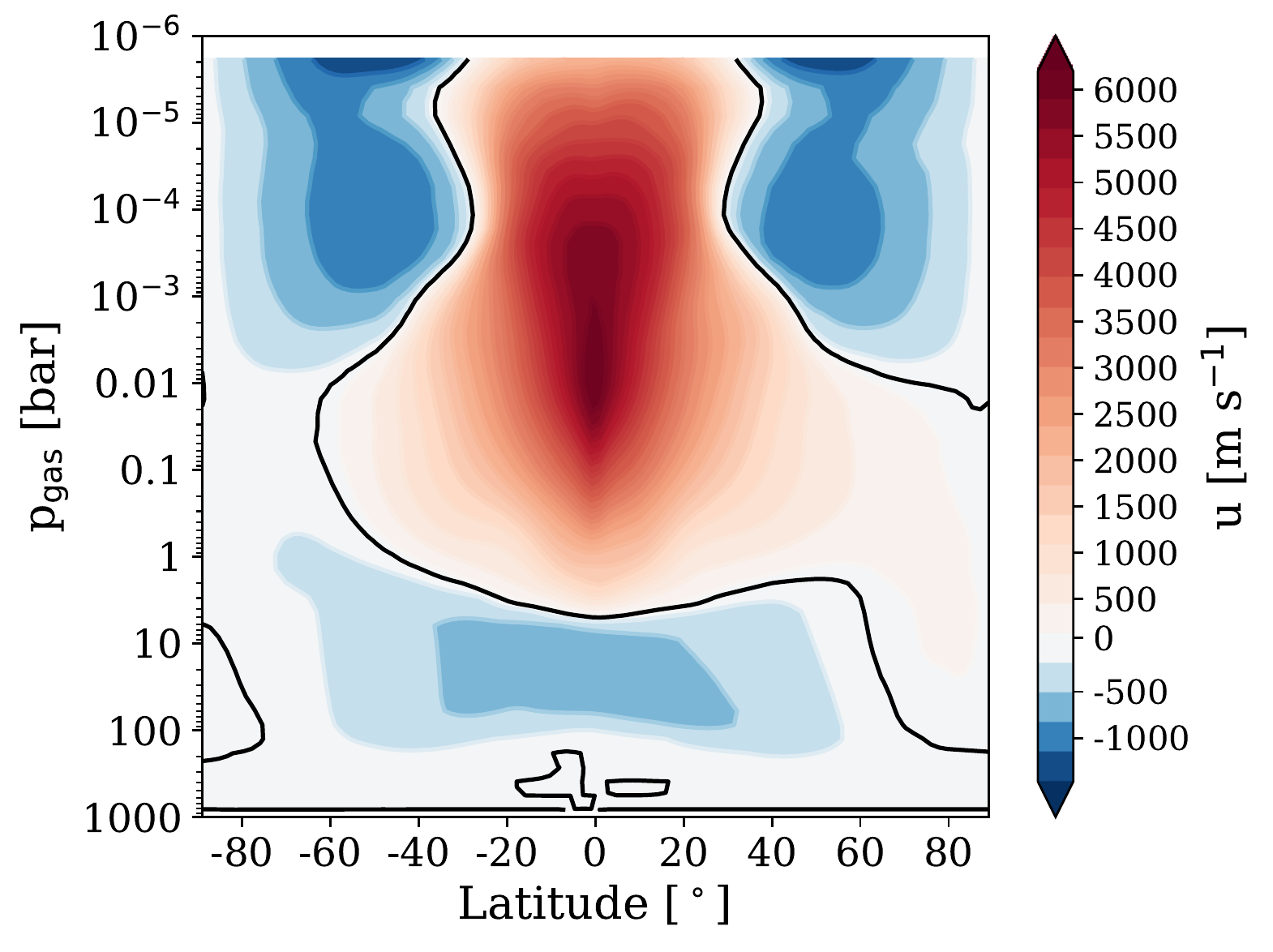}
   \includegraphics[width=0.49\textwidth]{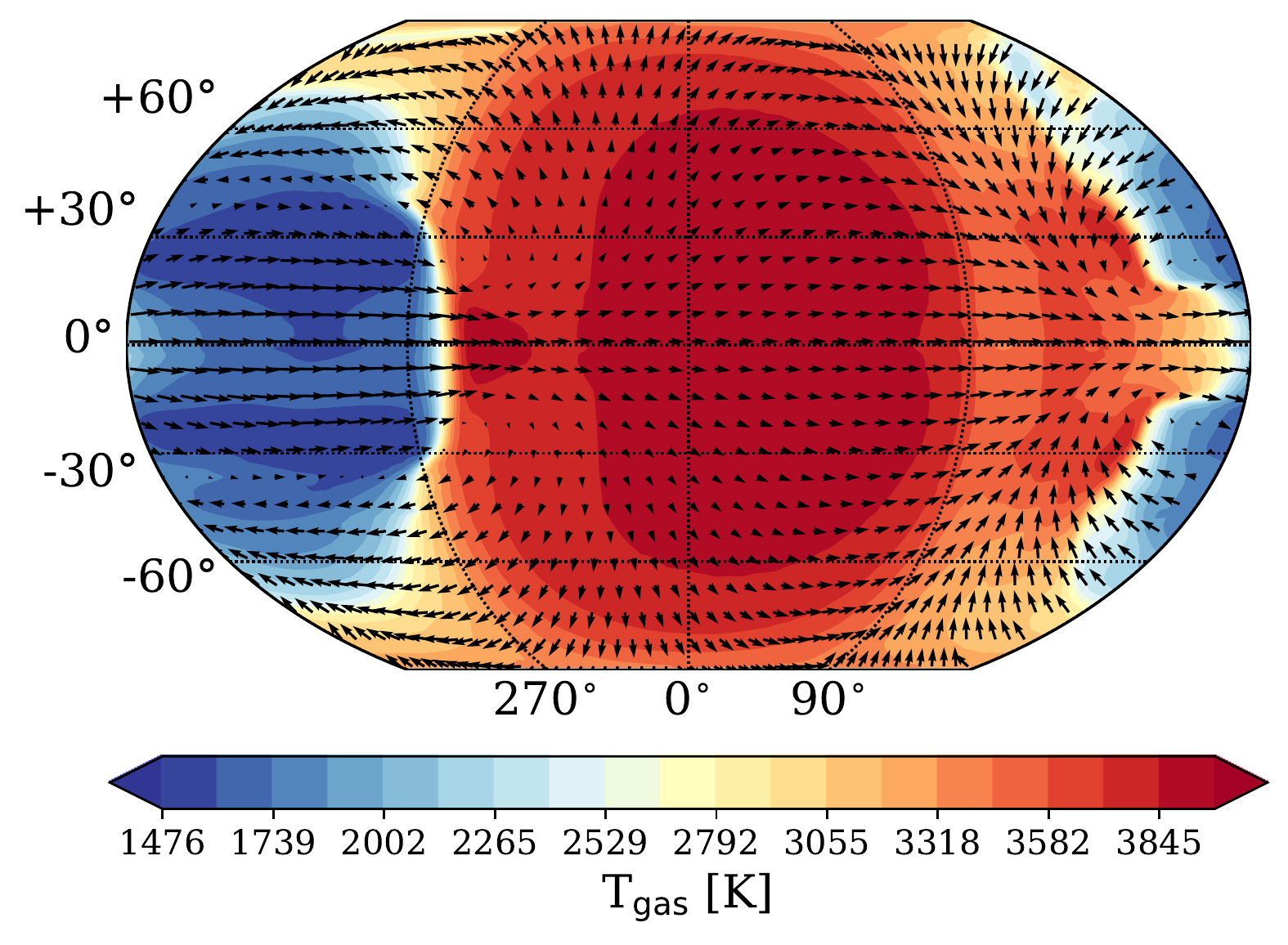}
   \includegraphics[width=0.49\textwidth]{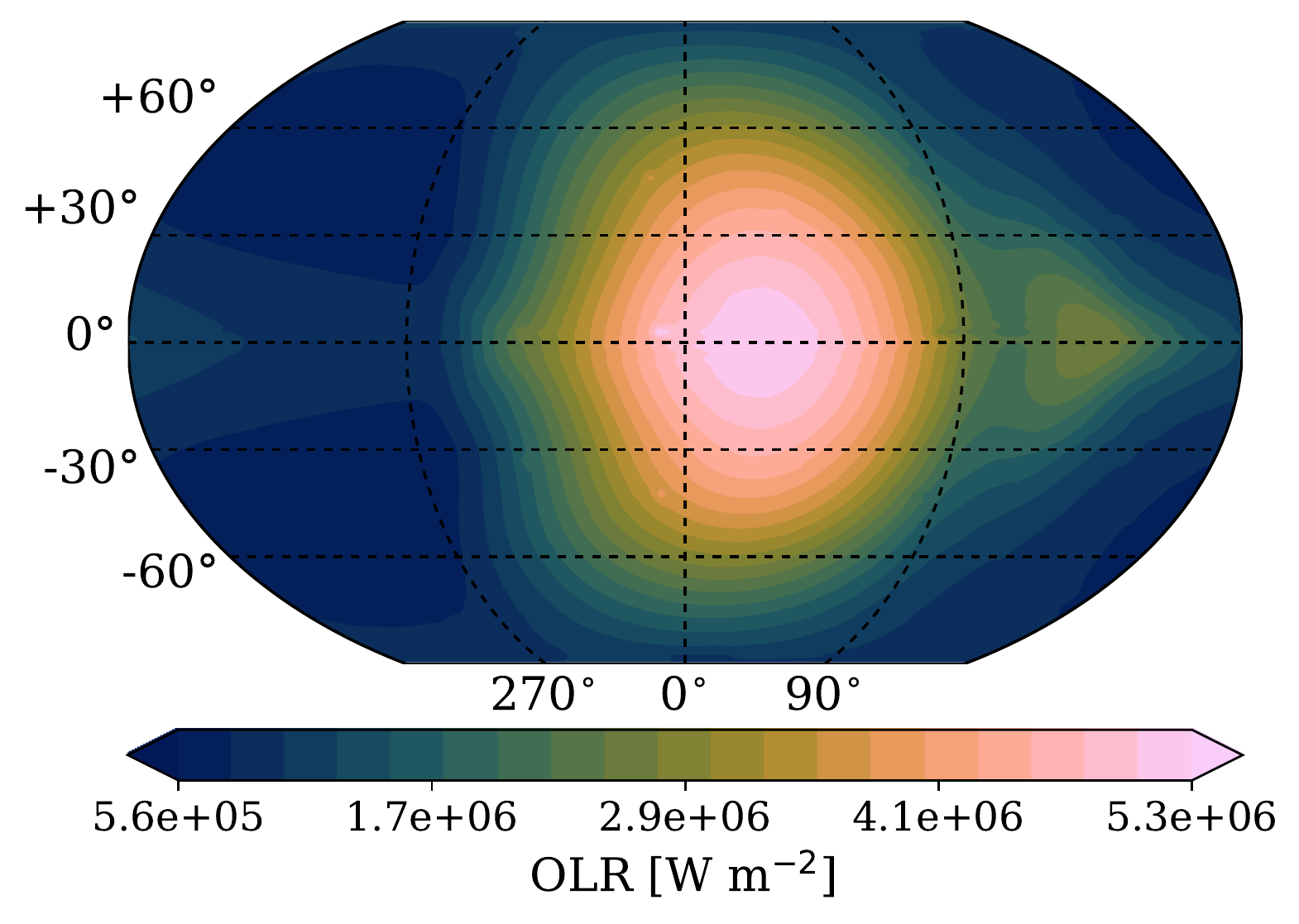}
   \caption{WASP-189b GCM results from Exo-FMS.
   Top left: Vertical T-p profiles at the equatorial regions (solid coloured lines) and polar column (dashed).
   Top right: Zonal mean zonal velocity.
   Bottom left: Temperature map at 1 mbar.
   Bottom right: Outgoing longwave radiation (OLR) map.}
   \label{fig:WASP189_GCM}
\end{figure*}

\subsection{Dynamic upper atmosphere}

\begin{figure*} 
   \centering
   \includegraphics[width=0.49\textwidth]{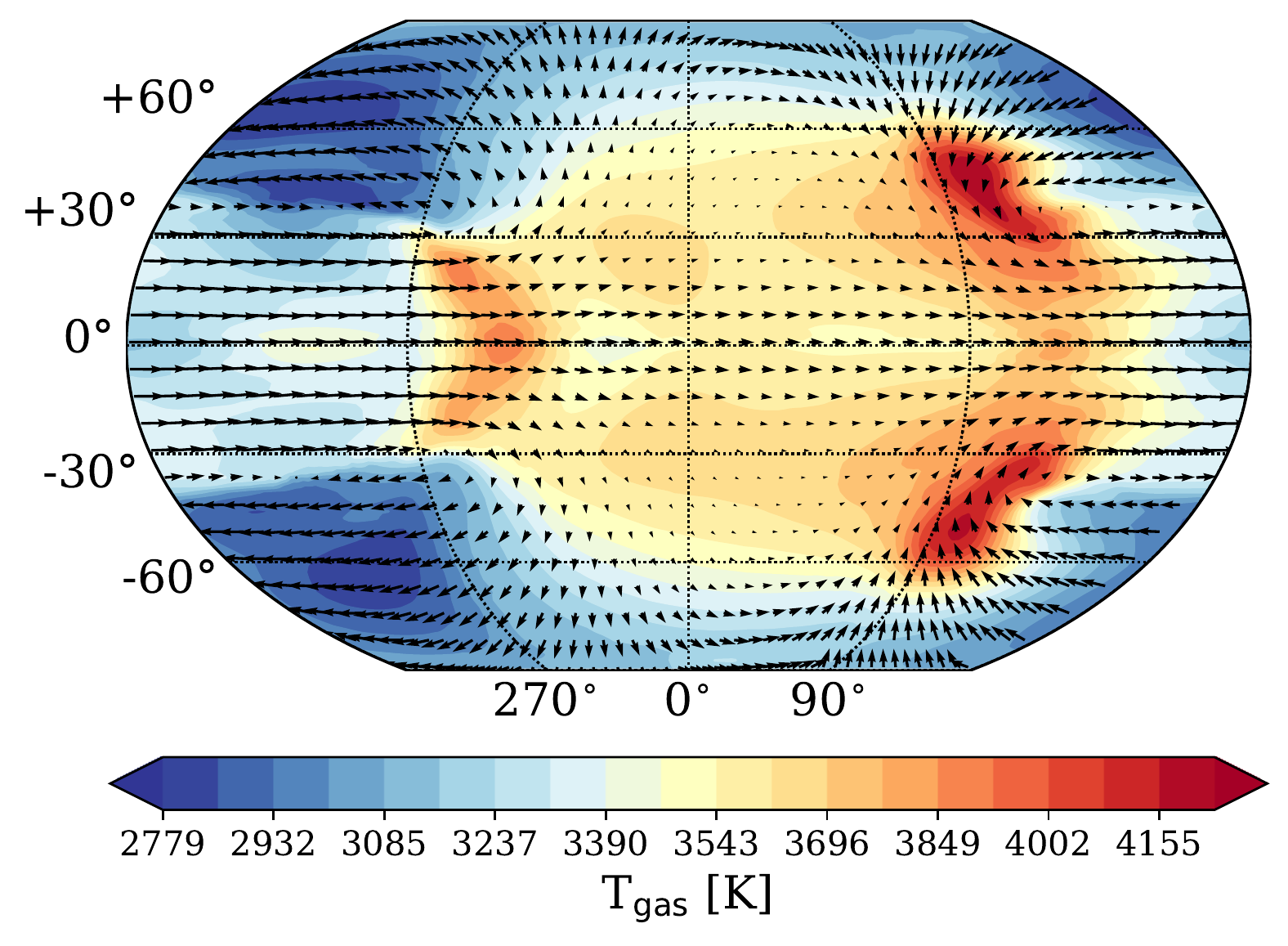}
   \includegraphics[width=0.49\textwidth]{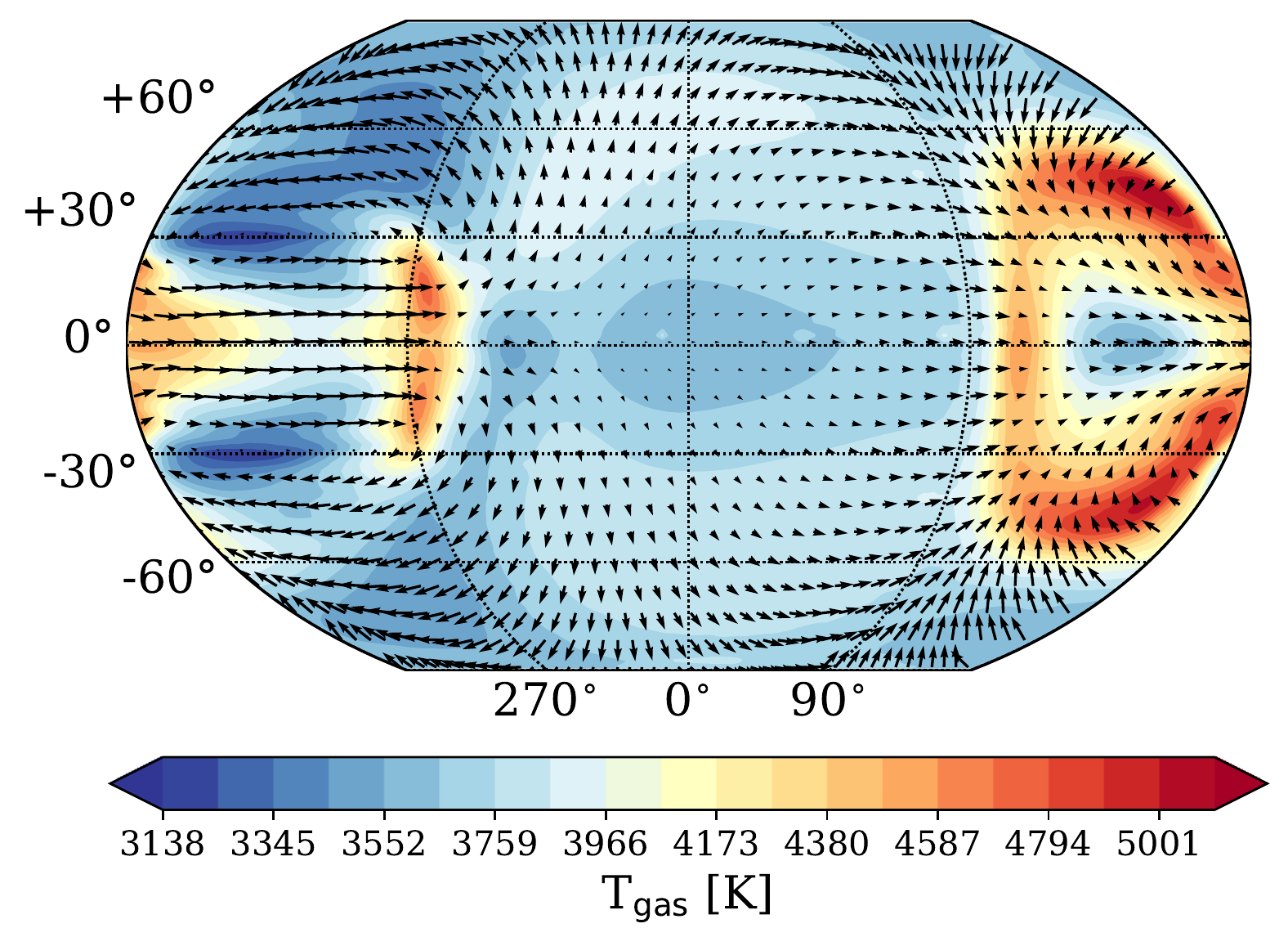}
   \caption{Lat-lon temperature map at 10$\mu$bar for WASP-121b (left) and WASP-189b (right) GCM models.}
   \label{fig:GCM_10ub}
\end{figure*}

Our GCM results show that a strong temperature inversions happen at very low pressures due to the absorption of UV irradiation.
In Figure \ref{fig:GCM_10ub} we show the temperature and wind direction quiver maps at 10$\mu$bar.
These show dynamical patterns that are unexpected at such low pressures and indicative of a separate and distinct wave behaviour occurring in these regions compared to the deeper jet forming layers.
This is also untypical as it is expected that the radiative-timescale should become very short at these upper atmosphere pressures, especially at these high temperatures, resulting in a rapid re-emission of absorbed shortwave radiation \citep[e.g.][]{Showman2009, Komacek2017}.
We therefore suggest that the inclusion of the strong UV-OPT absorbing species is responsible for the formation of this low-pressure region which is dynamically distinct from the deeper jet forming layers.
However, detailed analysis of the formation of these patterns is beyond the scope of the current study.

\section{Low resolution RT results}
\label{sec:low-res}

In this section, we post-process our WASP-121b and WASP-189b GCM results using the gCMCRT code \citep{Lee2022b} at low-resolutions.

\subsection{WASP-121b Low-resolution}

\subsubsection{Transmission spectrum}

\begin{figure*} 
   \centering
   \includegraphics[width=0.75\textwidth]{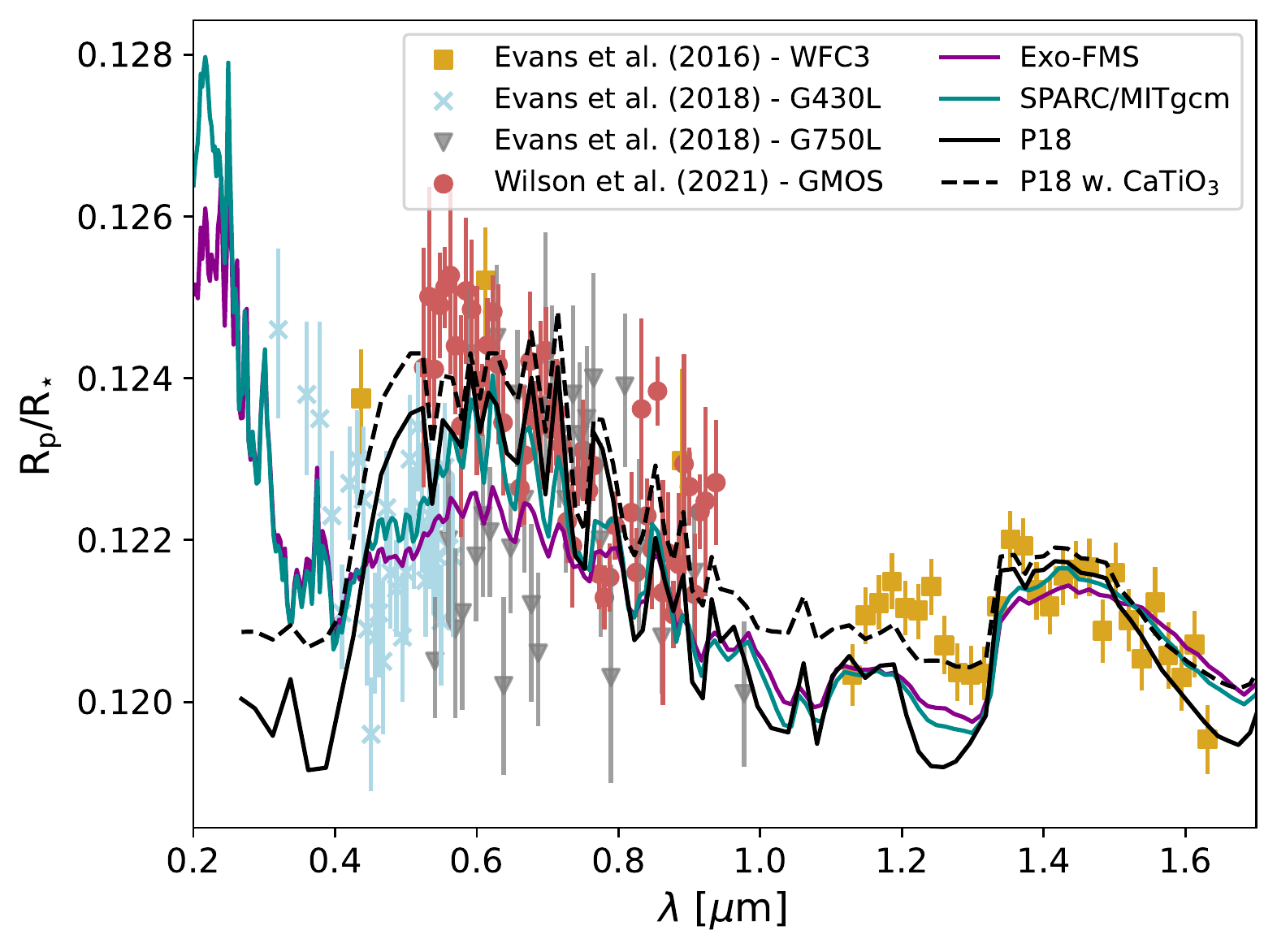}
   \caption{WASP-121b transmission spectrum post-processing for the Exo-FMS and SPARC/MITgcm models using gCMCRT as well as the processing performed in \citet{Parmentier2018} (P18).
   `P18 w. CaTiO$_{3}$' denotes the result from \citet{Parmentier2018} assuming CaTiO$_{3}$ cloud particles are present at the terminator regions.
   We compare to the \citet{Evans2016} and \citet{Evans2018} HST data and the \citet{Wilson2021} GMOS data.}
   \label{fig:WASP121_pp_lo_trans}
\end{figure*}

In Figure \ref{fig:WASP121_pp_lo_trans} we show the transmission spectrum post-processing results of our WASP-121b GCM model.
We assume that the atmosphere is cloud free.
We compare to the observational data of \citet{Evans2016, Evans2018} and \citet{Wilson2021}.
We find that the \citet{Parmentier2018} model better fits the variations and trends found in the observational data, suggesting that the SPARC/MITgcm run provides a better fit to the T-p conditions at the planetary terminators.
Differences are also seen between our 3D gCMCRT post-processing and the \citet{Parmentier2018} model at the HST WFC3 wavelengths, possibly indicating a difference between the processing methodologies.
Our gCMCRT post-processing also suggests that shortwards of 0.2$\mu$m, strong SiO absorption drastically increases the planetary radius, providing another species that could potentially fit the observational trends found in the \citet{Evans2018} data.

\subsubsection{Emission spectrum}

\begin{figure*} 
   \centering
   \includegraphics[width=0.49\textwidth]{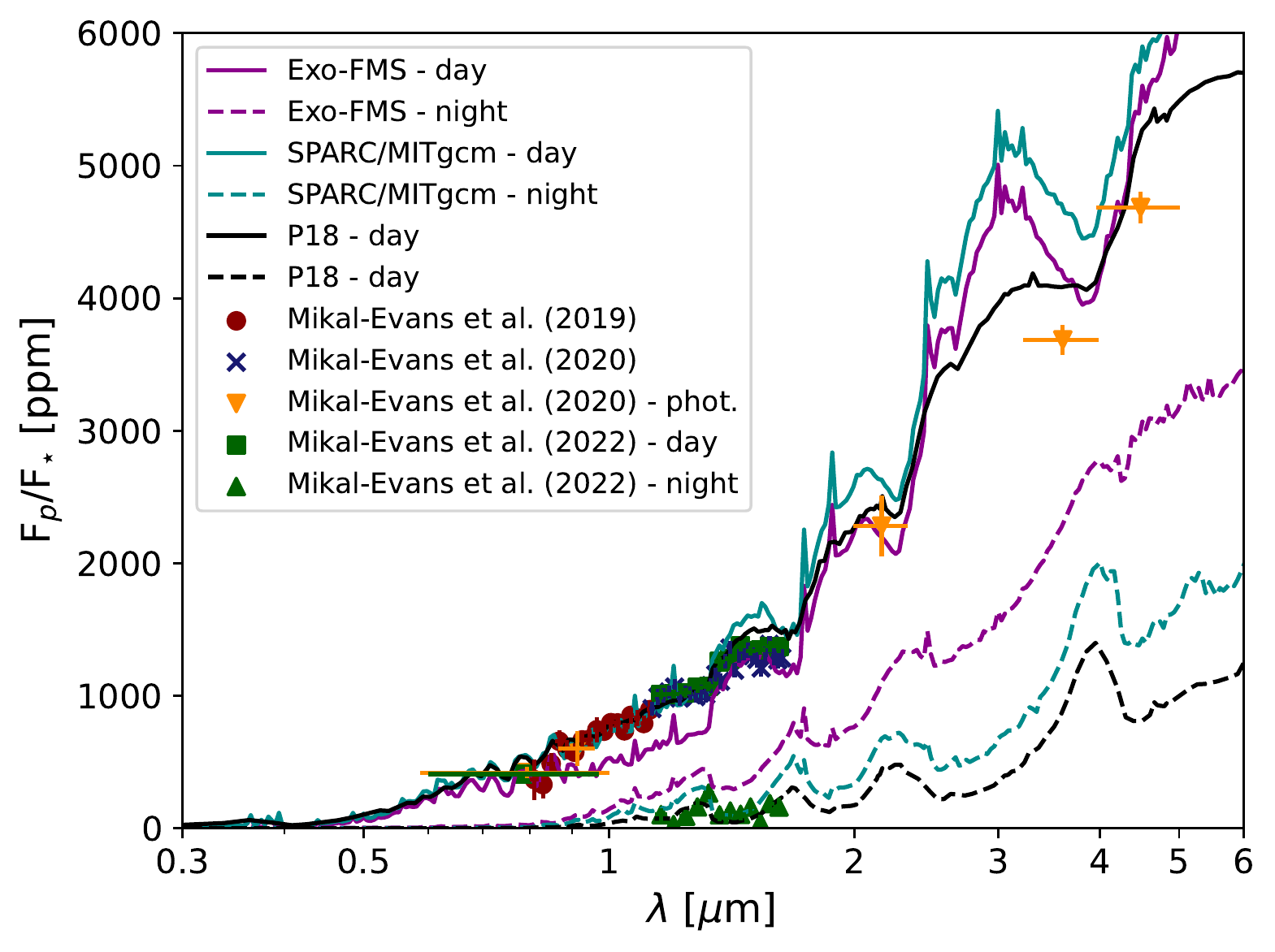}
   \includegraphics[width=0.49\textwidth]{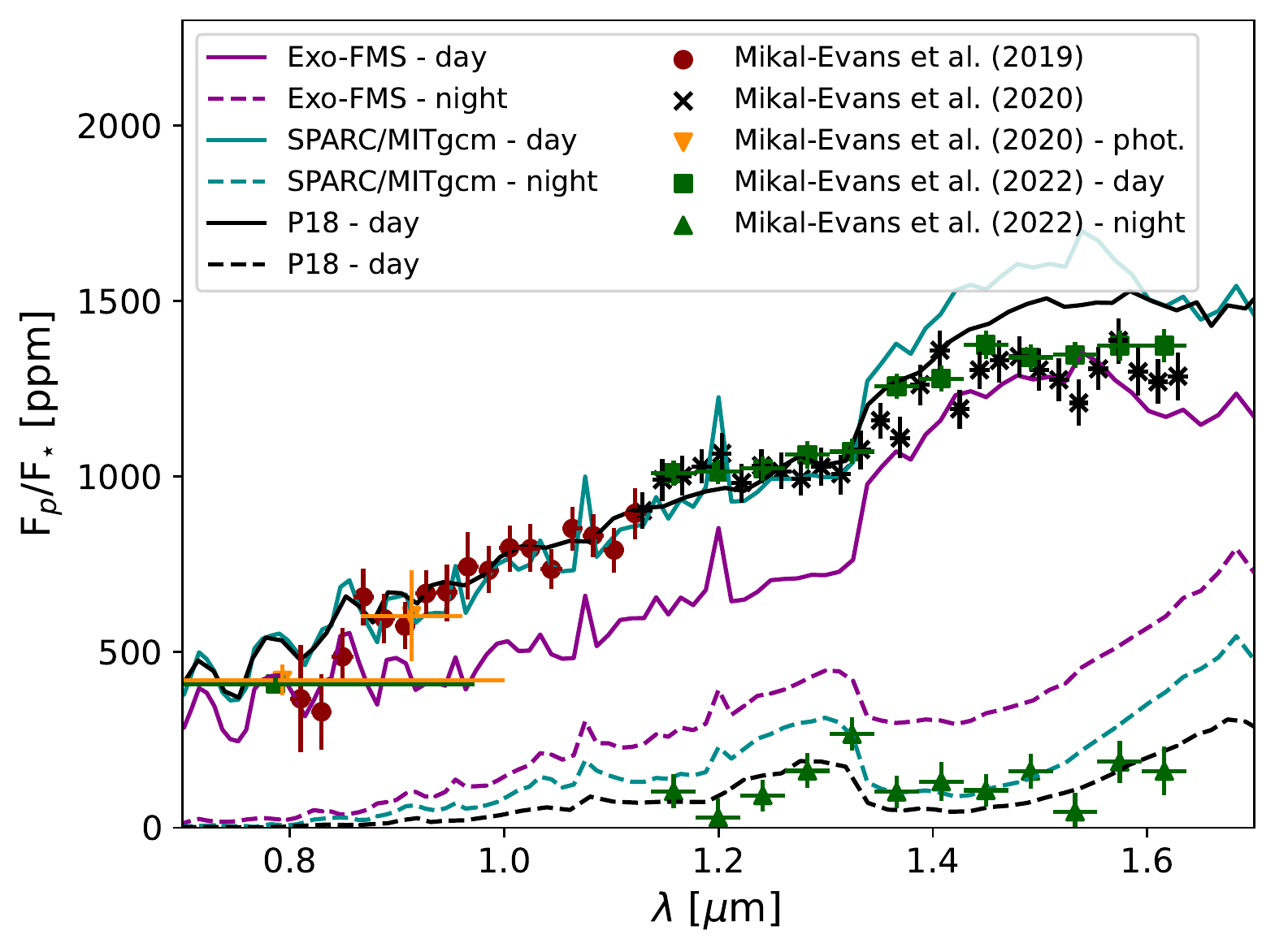}
   \caption{WASP-121b dayside and nightside emission spectra post-processing for the Exo-FMS and SPARC/MITgcm models.
    Right: Comparing to the \citet{Evans2019, Evans2020} and \citet{Mikal-Evans2022} dayside and nightside data.
    Left: Zoom into the HST data wavelength range.
    `P18' corresponds to the original \citet{Parmentier2018} post-processing.}
   \label{fig:WASP121_pp_lo_em}
\end{figure*}

In Figure \ref{fig:WASP121_pp_lo_em} we present the post-processing results of the WASP-121b model in emission.
Our processing of the \citet{Parmentier2018} GCM model produces a better fit than the Exo-FMS GCM for the HST WFC3 data from \citet{Evans2019,Evans2020}.
However, the Exo-FMS produces a better fit to the Spitzer photometric bands.
This suggests that both GCMs capture decently the T-p structure of the dayside atmosphere, with \citet{Parmentier2018} fitting well the deeper atmosphere while Exo-FMS fitting the upper atmosphere.
Figure \ref{fig:WASP121_pp_lo_em} also shows clearly the differences in the day-night transport efficiency between the models, with the SPARC/MITgcm model showing a reduced flux emanating from the nightside compared to the Exo-FMS GCM.

\subsubsection{TESS phase curve}

\begin{figure*} 
   \centering
   \includegraphics[width=0.49\textwidth]{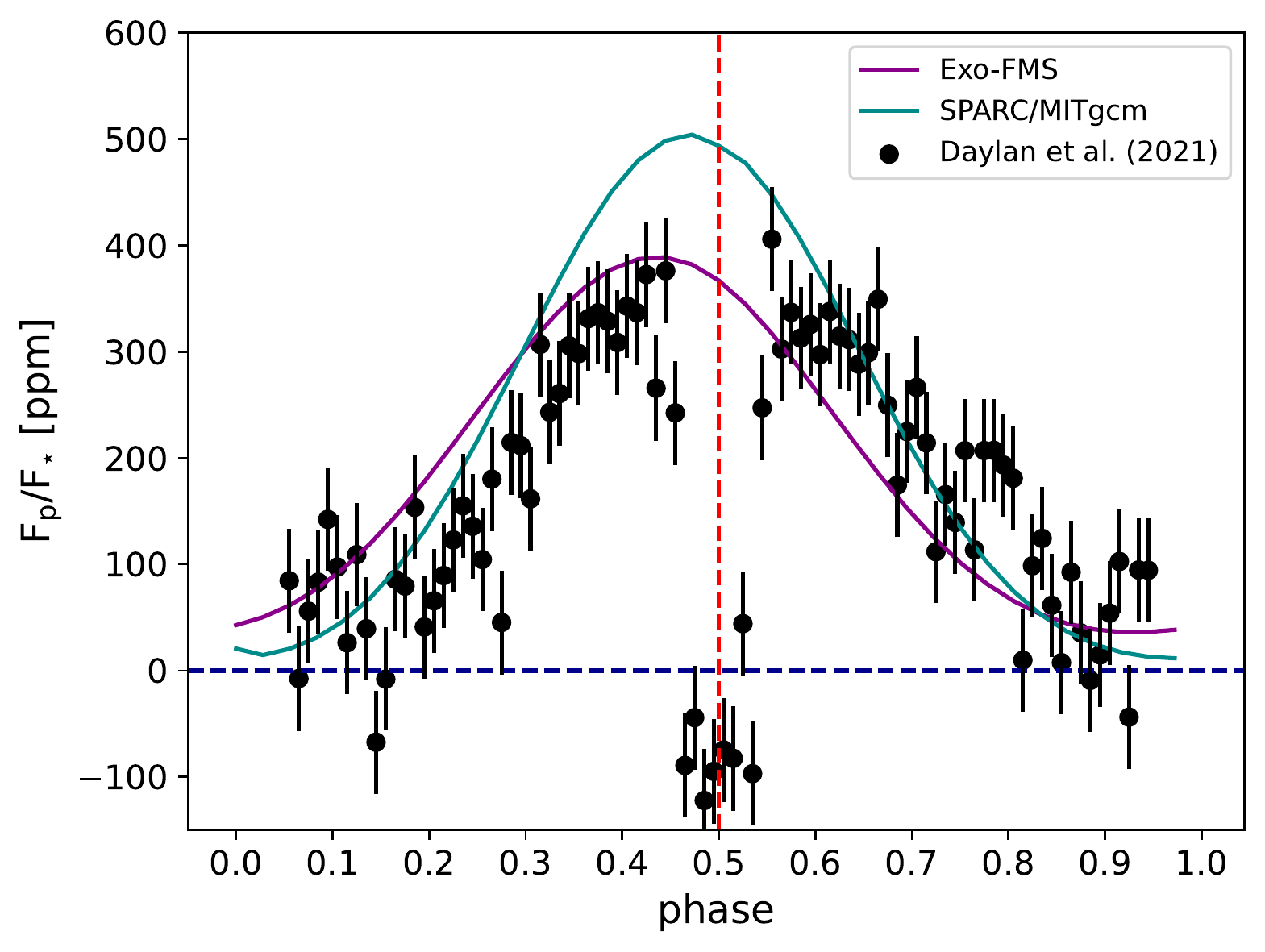}
   \caption{WASP-121b TESS phase curves from \citet{Daylan2021} (points) compared to the SPARC/MITgcm and
   Exo-FMS post-processing model.}
   \label{fig:WASP121_pp_lo_TESS}
\end{figure*}

Figure \ref{fig:WASP121_pp_lo_TESS} shows the synthetic TESS phase curves from both WASP-121b GCM models compared to the data presented in \citet{Daylan2021}.
Generally, the \citet{Parmentier2018} model shows a better fit to the phase offset, while the Exo-FMS model is more in line with the emitted flux values, but showing too much of an eastward shift compared to the TESS data.
However, the Exo-FMS shows a better fit to the day/night contrast in the TESS band.

\subsection{WASP-189b Low-resolution}

In this section, we show the low resolution post-processing results for the WASP-189b GCM model.

\subsubsection{Transmission and emission spectrum}

\begin{figure*} 
   \centering
   \includegraphics[width=0.49\textwidth]{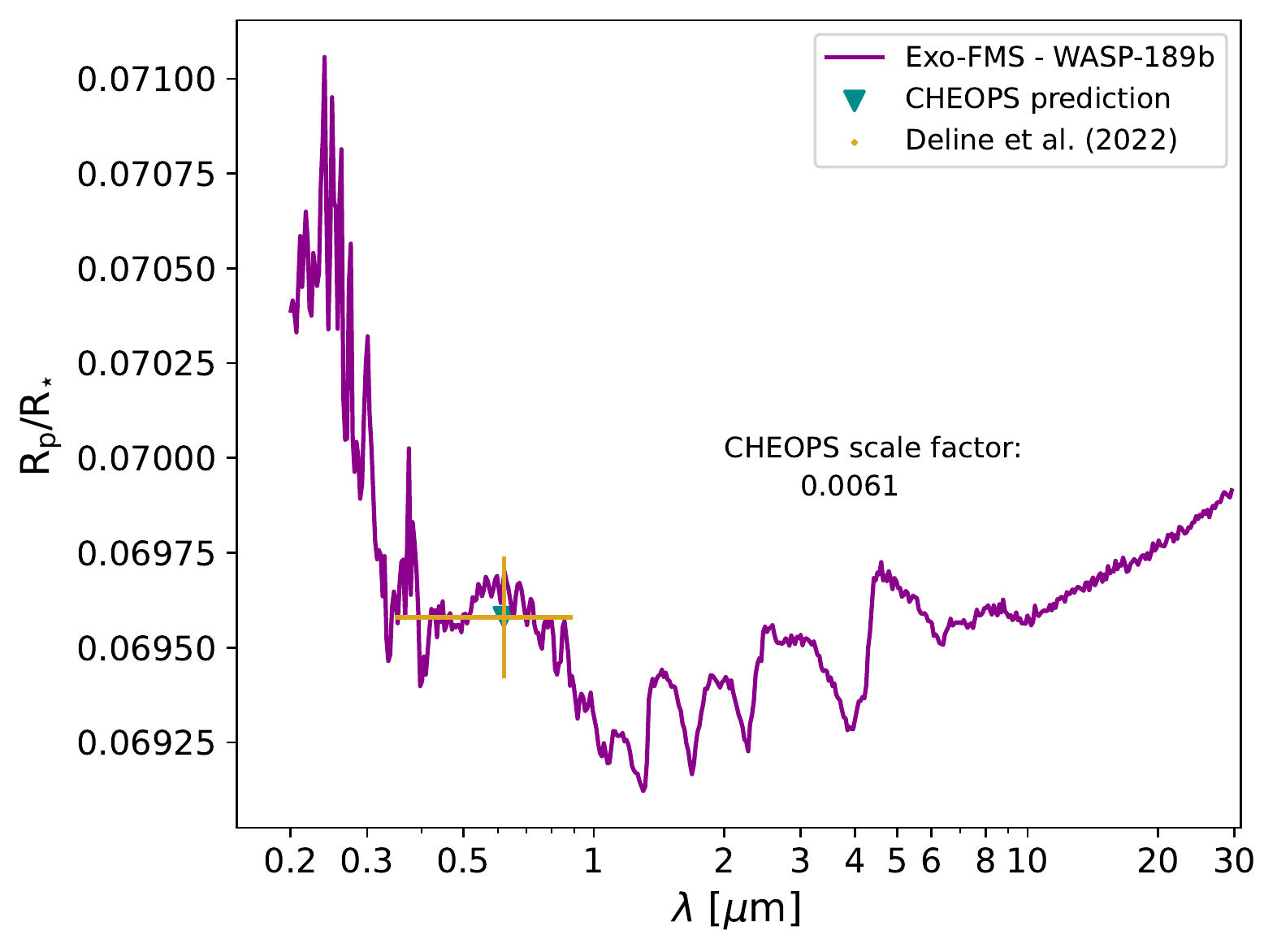}
   \includegraphics[width=0.49\textwidth]{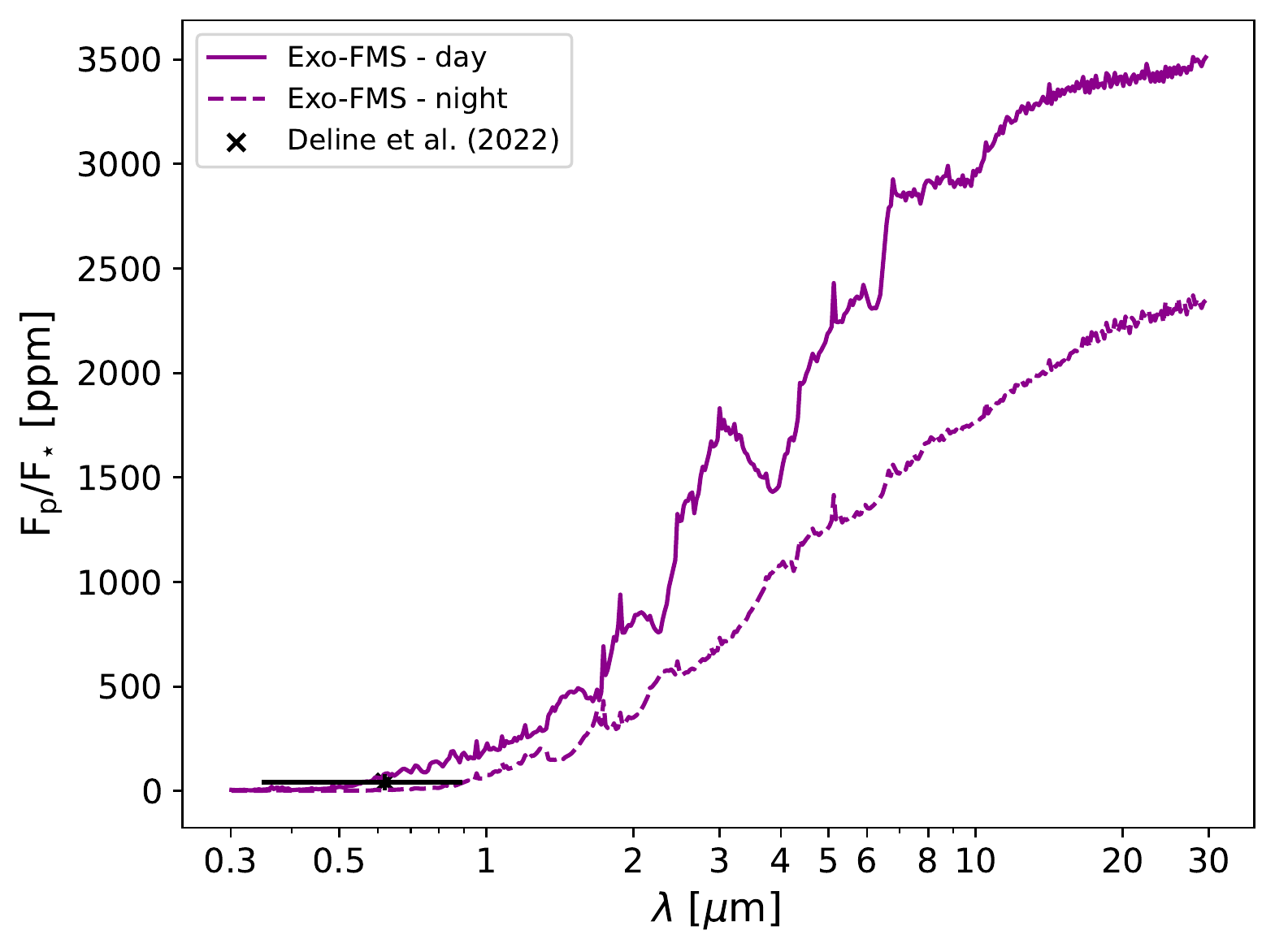}
   \caption{WASP-189b Exo-FMS post-processing for the transmission spectrum (left) and dayside and nightside emission spectrum (right).
   The model is compared to the \citet{Deline2022} CHEOPS data.
   A vertical shift of the model transmission spectrum is applied to best fit the CHEOPS datapoint.}
   \label{fig:WASP189_pp_lo}
\end{figure*}

In Figure \ref{fig:WASP189_pp_lo} we present the transmission and emission spectra post-processing of the Exo-FMS GCM at low resolution.
Again, we assume the atmosphere is cloud free.
Our WASP-189b model shows similar trends to the WASP-121b model in both transmission and emission.
Similar to the WASP-121b model, the absorption of SiO at NUV wavelengths ($<$ 0.3 $\mu$m) increases the transit depth dramatically.
So far only CHEOPS data from \citet{Deline2022} is available, which we add to the spectra plots.
We use the R$_{\rm p}$/R$_{\star}$ value from \citet{Deline2022} to scale the model transmission spectra to the observed value.
We produce a CHEOPS 'prediction' with the model spectrum, then match this prediction point with the CHEOPS point.
This gives a scale factor of 0.0061 R$_{\rm p}$/R$_{\star}$ between the non-scaled model and the CHEOPS point.

\subsubsection{CHEOPS phase curve}

\begin{figure*} 
   \centering
   \includegraphics[width=0.49\textwidth]{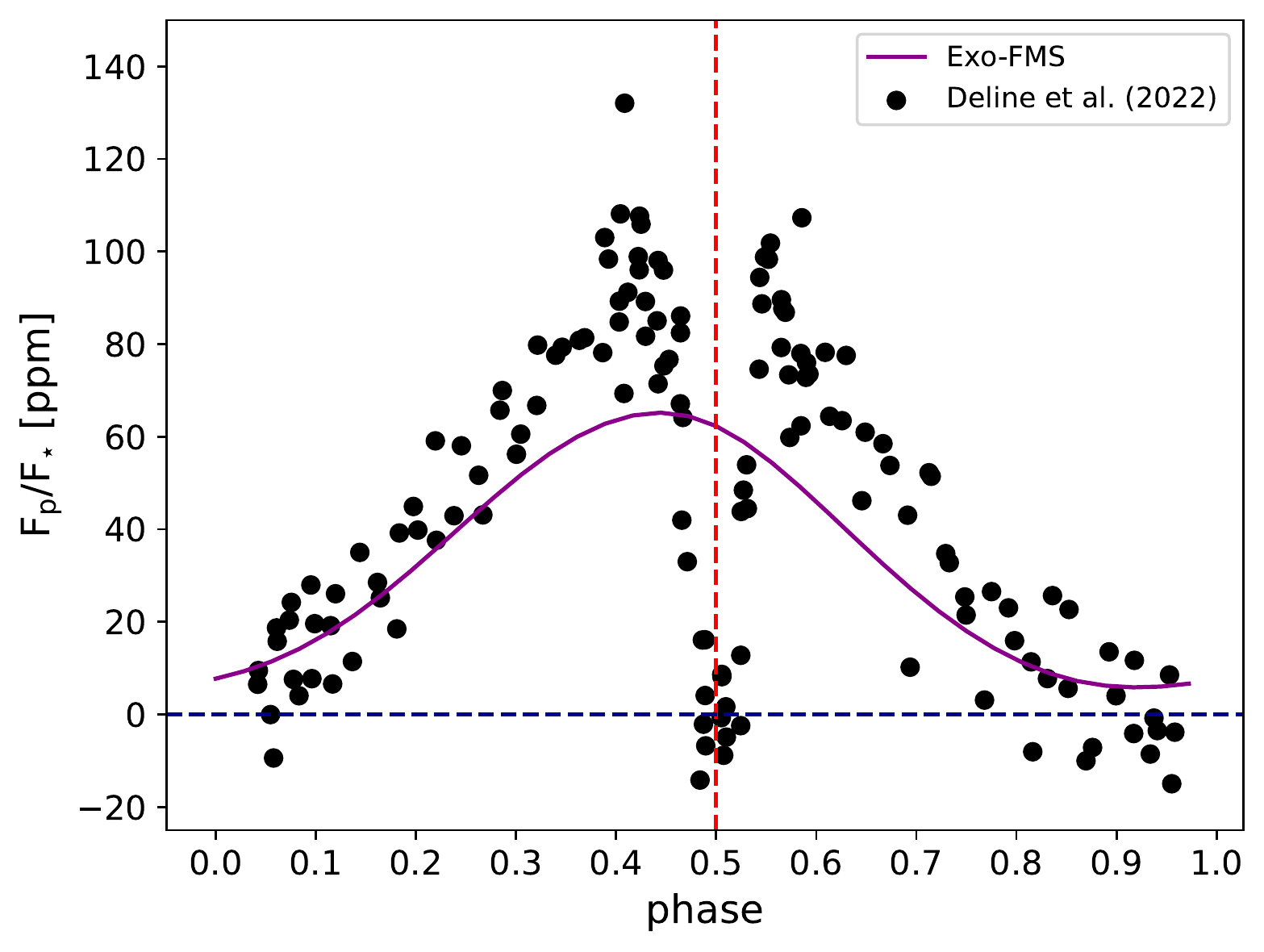}
   \caption{WASP-189b Exo-FMS GCM post-processing comparing to the CHEOPS phase curve data presented in \citet{Deline2022}.}
   \label{fig:WASP189_pp_lo_CHEOPS}
\end{figure*}

In Figure \ref{fig:WASP189_pp_lo_CHEOPS} we show the Exo-FMS GCM model processing results compared to the CHEOPS phase curve data in \citet{Deline2022}.
This plot suggests that the dayside of the GCM is too cool at the pressures where the CHEOPS bandpass probes.
However, the GCM provides a reasonable fit to the offset in this case compared to the WASP-121b TESS phase curve data.
The GCM also fits well the emission from the dayside of planet, suggesting that the nightside T-p structure in  is being reasonably captured by the GCM for the CHEOPS bandpass.

\section{High resolution RT results}
\label{sec:high-res}

In this section, we post-process the SPARC/MITgcm WASP-121b model and the Exo-FMS WASP-189b model at high spectral resolution to produce synthetic observational data and then cross-correlate with molecular and atomic spectral templates to produce synthetic molecular/atomic species detections.
We chose the SPARC/MITgcm GCM model for WASP-121b since its low-resolution post-processing results from Section \ref{sec:low-res} fit the data better than the Exo-FMS model.
This is especially evident in the hot spot shift seen in the TESS phase curve and the day/night contrast in the processed emission spectra, which suggests that the SPARC/MITgcm velocity fields are more in line with the real object compared to the Exo-FMS.

\subsection{WASP-121b High-resolution}

\begin{figure*}
    \centering
    \includegraphics[width=\textwidth]{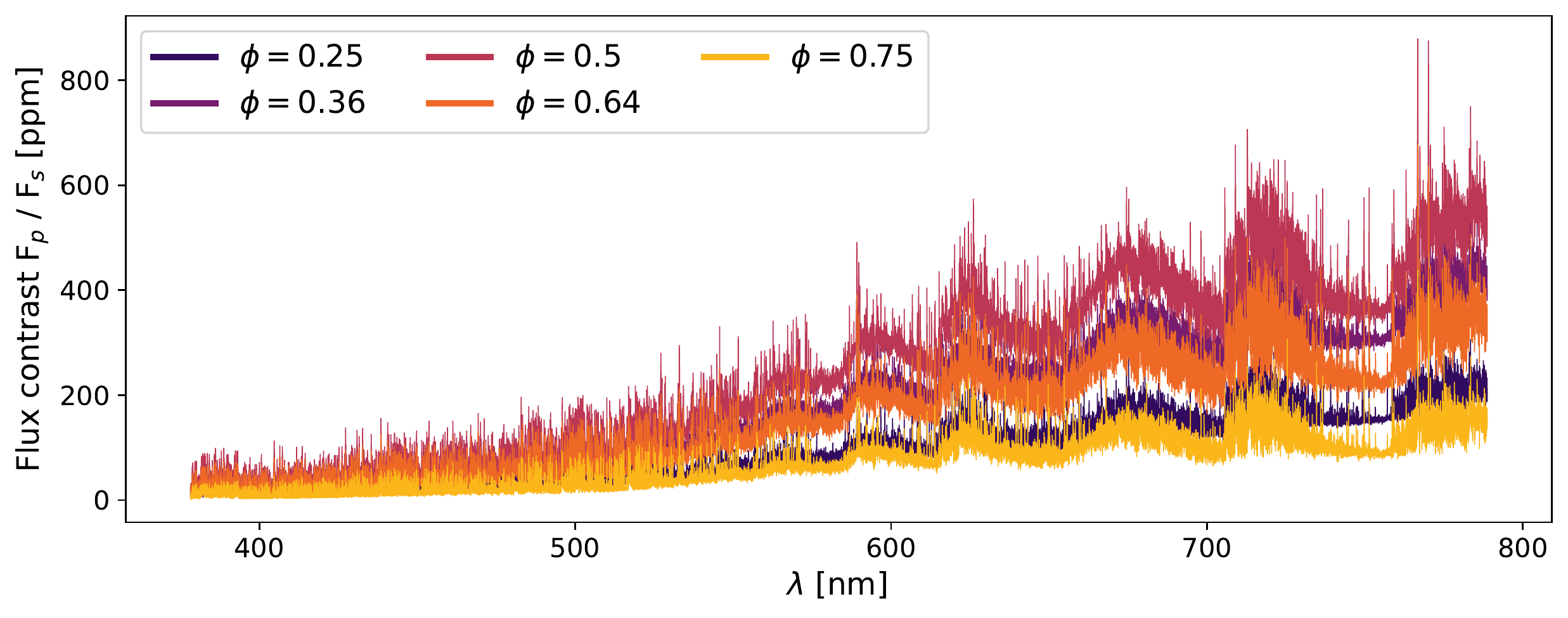}
    \caption{Modelled high-resolution planet-to-star flux ratio of WASP-121b at phases $\phi = 0.25, 0.36, 0.5, 0.64$ and $0.75$. The secondary eclipse occurs at $\phi = 0.5$, where the emission spectrum is the strongest. $\phi < 0.5$ indicate the east-wards spectra prior to secondary eclipse, and $\phi > 0.5$ indicate the west-wards spectra after secondary eclipse. The star is modelled as a blackbody at temperature $T_{\rm eff}=6459\pm140$ \,K \citep{Delrez2016} and stellar radius of $R_\ast = 1.458 R_\odot$ \citep{Bourrier2020}}.
    \label{fig:W121-em-spec}
\end{figure*}

\begin{figure*} 
   \centering
   \includegraphics[width=0.99\textwidth]{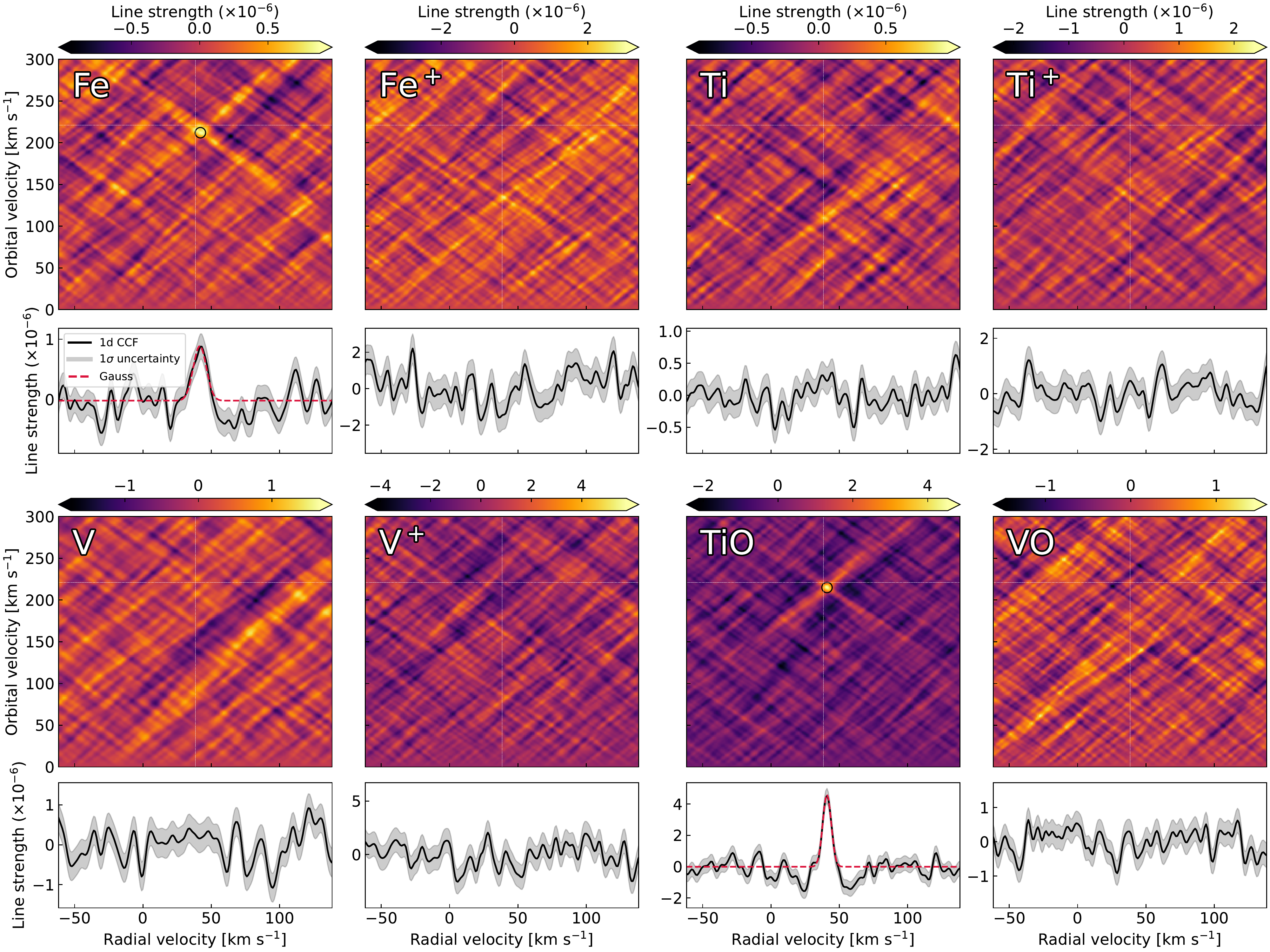}
   \caption{Synthetic cross-correlation results for the WASP-121b GCM processing. Rows 1\&3: Velocity–velocity diagrams showing the modeled emission signals of Fe, Fe$^{+}$, Ti, Ti$^{+}$, V, V$^{+}$, TiO, and VO in the rest frame of the star. The horizontal dashed lines indicate the orbital velocity at which the signals are expected to peak $\left(v_{\rm orb} = \frac{2 \pi a}{P} = 221.1\,{\rm km s}^{-1}\right)$. The vertical dashed lines indicate the expected systemic velocity of $v_{\rm sys} = 38.3\,{\rm km s}^{-1}$; see \citet{Bourrier2020} for orbital parameters. Black circles indicate the peak position where the signal is extracted. If no signal is detected, the signal is extracted at the true orbital velocity.  Rows 2\&4: Cross-correlation functions stacked in the rest frame of the planet at the extracted orbital velocity. The shaded regions indicate the 1$\sigma$ uncertainties. Dashed lines show the best-fit Gaussian model.}
   \label{fig:WASP121_pp_hi}
\end{figure*}

\subsubsection{Emission spectra}

In Figure \ref{fig:W121-em-spec}, we show the emission spectra after processing the SPARC/MITgcm of WASP-121b at high spectral resolution. For our high-resolution post-processing of WASP-121b, we create synthetic observations following the observational strategy of Hoeijmakers et al. (in prep). They observed WASP-121b with the echelle spectrograph ESPRESSO on ESO's Very Large Telescope in Chile \citep{pepe_espressovlt_2021} (R = 140,000) over phases from 0.25-0.75, not including the secondary eclipse of the planet. We produce emission spectra over this phase range containing the species Fe, Fe$^{+}$, H$_{2}$O, K, Na, OH, SiO, Ti, Ti$^{+}$, TiO, V, V$^{+}$ and VO between wavelengths 377 - 790\,nm. The contrast of the spectra varies over the course of the phase coverage, with peak contrast at secondary eclipse. The spectra are not symmetric around secondary eclipse. We further produce cross-correlation templates for Fe, Fe$^{+}$, Ti, Ti$^{+}$, TiO, V, V$^{+}$ and VO at phase $\phi = 0.5$ using the GCM output.
We use templates produced from the GCM model since it was found in \citet{Beltz2021} that using 3D GCM templates increases the detection significance of molecules compared to 1D templates.

\subsubsection{Synthetic observations}

To create synthetic observations, we generate spectra by adding the spectral flux of the modelled planet to the stellar spectrum. The stellar spectrum is retrieved from the PHOENIX database \citep{Husser2013}, assuming $\log{g}=4.24^{+0.011}_{-0.012}$, $T_{\rm eff}=6459\pm140$ \,K \citep{Delrez2016} and broadening to a projected rotational velocity of $v\sin{i)} = 13.6$ \kmpers \citep[][]{Bourrier2020}. To add the spectra, we correct for their Keplerian velocities at each phase and assume a contrast of
\begin{equation}
    \left(\frac{R_{\rm p}}{R_\star}\right)^2 = 0.01526,
\end{equation}
with $\frac{R_{\rm p}}{R_\star}$ = 0.12355 \citep[][]{Bourrier2020}, resulting in

\begin{equation}
    S = F_s + F_p = F_s \left(1+ \frac{F_p}{F_s}\right) = f_s R_s^2 \left(1 + \left(\frac{R_{\rm p}}{R_\star}\right)^2 \frac{f_p}{f_s}\right).
\end{equation}

The combined spectra are then decomposed into the orders corresponding to the echelle orders of ESPRESSO and multiplied by the blaze function specific to the instrument.
In a final step, Gaussian noise with a signal-to-noise ratio (SNR) based on the true SNR of the observations of Hoeijmakers et al. (in prep.). For each order, we calculate the true SNR over a 20 pixels running standard deviation window. The spectra are not corrected for Earth's barycentre velocity and no telluric lines are added, because this would only lead to a re-interpolation of the spectra as well as a dividing out of the telluric lines.

\subsubsection{Cross-correlation analysis}

We perform cross-correlations\footnote{The computer code for performing cross correlations is publicly available at \url{https://github.com/hoeijmakers/tayph/}. Documentation, instructions and a data demonstration can be found at \url{https://tayph.readthedocs.io}.} with templates for Fe, Fe$^{+}$, Ti, Ti$^{+}$, TiO, V, V$^{+}$ and VO based on the GCM models, presented in Appendix \ref{fig:WASP-121b_templates}.
We follow the methodology in \citet{Prinoth2022}, except for the Doppler shadow correction and vertical detrending, which is also caused by aliases of the Doppler shadow. We do not introduce the Rossiter-McLaughlin effect, so correction is not required. We produce \kpvsys diagrams with radial velocities from -1000 to 1000 \kmpers and orbital velocities from 0 to 300 \kmpers in steps of 1 \kmpers. The combined results of the cross-correlation with eight epochs (four before and four after the secondary eclipse) are shown in Figure \ref{fig:WASP121_pp_hi}. The cross-correlation results show detections for Fe and TiO. It is particularly noteworthy that TiO is detected with high significance, which contrasts with the non-detection of TiO in the transmission spectrum of WASP-121b in \citet{Hoeijmakers2020}.
We also do not detect atomic V which was detected in \citet{Hoeijmakers2020}.
However, the dayside emission spectra properties of a planet are very different from the transmission spectra and probe different regions.
Hoeijmakers et al. (in prep) will examine the dayside spectrum of WASP-121 at high resolution and will be able to discern if our predictions here hold for the dayside or are in line with the transmission data.

\subsection{WASP-189b High-resolution}

\begin{figure*}
    \centering
    \includegraphics[width=\textwidth]{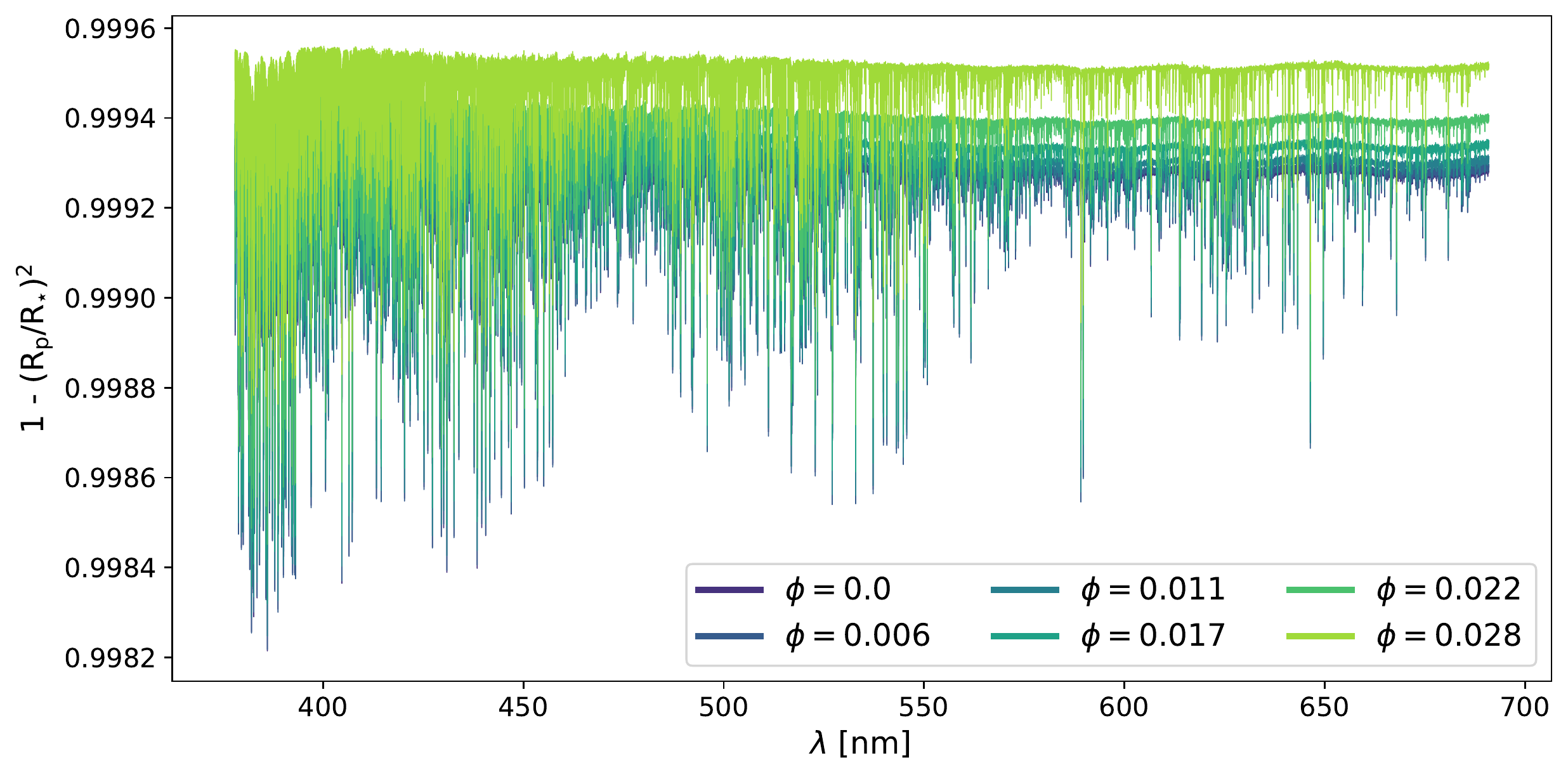}
    \caption{Modelled transmission spectrum in $1- \delta^2$ of WASP-189b at phases $\phi = 0, 0.006, 0.011, 0.01, 0.022\,{\text{and}}\,0.028$. The transmission spectrum is symmetric around phase $\phi =0$, which corresponds to the transit midpoint. The transmission spectrum is expected to be stronger towards the centre of the transit due to the reduced effect of limb-darkening at mid-transit. Note that the factor $\alpha = 3$ is not included in the presentation of this plot.}
    \label{fig:W189-trans-spec}
\end{figure*}

\begin{figure*} 
   \centering
   \includegraphics[width=0.99\textwidth]{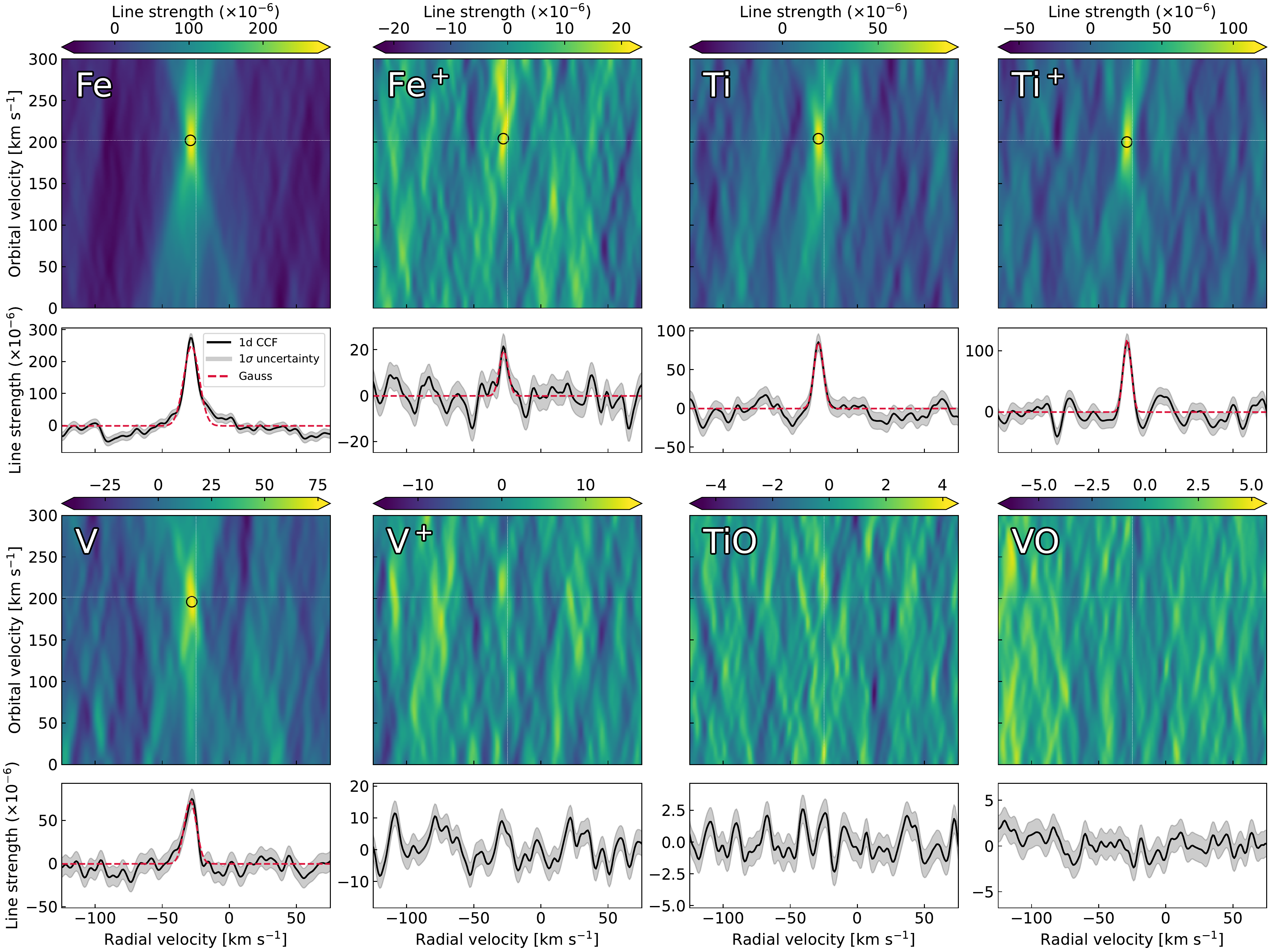}
   \caption{Synthetic cross-correlation results for the WASP-189b GCM model processing. Rows 1\&3: Velocity–velocity diagrams showing the modeled emission signals of Fe, Fe$^{+}$, Ti, Ti$^{+}$, V, V$^{+}$, TiO, and VO in the rest frame of the star. The horizontal dashed lines indicate the orbital velocity at which the signals are expected to peak ($v_{\rm orb} = \frac{2 \pi a}{P} = 201.8\,{\rm km s}^{-1}$). The vertical dashed lines indicate the expected systemic velocity of $v_{\rm sys} = -24.8\,{\rm km s}^{-1}$; see \citet{Lendl2020} and \citet{Anderson2018} for orbital parameters. Black circles indicate the peak position where the signal is extracted. If no signal is detected, the signal is extracted at the true orbital velocity. Rows 2\&4: Cross-correlation functions stacked in the rest frame of the planet at the extracted orbital velocity. The shaded regions indicate the 1$\sigma$ uncertainties. Dashed lines show the best-fit Gaussian model.}
   \label{fig:WASP189_pp_hi}
\end{figure*}

\subsubsection{Transmission spectra}

In Figure \ref{fig:W189-trans-spec}, we show the transmission spectra after processing the Exo-FMS GCM of WASP-189b at high spectral resolution. For our high-resolution study of WASP-189b, we follow the observational strategy of \citet{Prinoth2022} and generate transmission spectra over the transit phases with the HARPS(N) resolution of R\,$\sim$\,120,000, between wavelengths 383 and 693\,nm. We construct a total opacity model that includes Fe, Fe$^{+}$, H$_{2}$O, K, Na, OH, Ti, Ti$^{+}$, TiO, V, V$^{+}$ and VO line opacities. The combined total opacities are then used in the calculations of Doppler-shifted and rotationally-broadened spectra for 11 phases between 0.972 and 0.028. Non-doppler-shifted individual cross-correlation templates for each species are also created assuming that the planet is at mid-transit, and are shown in Appendix \ref{fig:WASP-189b_templates}.

\subsubsection{Synthetic observations}

To create synthetic observations, we use the HARPS and HARPS-N observations analysed in \citet{Prinoth2022} to determine the true SNR. The stellar spectrum of WASP-189 is obtained from the PHOENIX database  assuming $\log{g}=3.8 \pm 0.2$, $T_{\rm eff}=8000 \pm 80$ \,K \citep{Lendl2020}. It is then rotationally broadened with a projected rotational velocity of $v\sin{i)} = 93.1$ \kmpers \citep{Lendl2020}. After correcting for the Keplerian velocities, the combined star-planet spectrum can be calculated as
\begin{equation}
    S = F_s T = F_s \left(1 - \alpha \delta \right)
    \label{eq:W189_trans}
\end{equation}
where $\delta$ is the transit depth. The factor of $\alpha = 3$ in Eq. \eqref{eq:W189_trans} serves as a tuning factor to account for potential uncertainty of the limb-darkening properties of the host star.
\citet{Morello2017} suggest that the Claret 4 parameter law \citep{Claret2000} is more accurate at modelling limb darkening behaviour compared to the simple quadratic law we use in this study, especially at optical wavelengths. The combined star-planet spectrum is decomposed into the echelle orders of the HARPS and HARPS-N spectrographs, multiplied by the blaze functions of the instruments, and Gaussian noise is applied based on the calculated true SNR.

\subsubsection{Cross-correlation analysis}

We perform cross-correlations with templates for Fe, Fe$^{+}$, Ti, Ti$^{+}$, TiO, V, V$^{+}$ and VO based on the GCM models. The combined cross-correlation results for all five observation nights are shown in Figure \ref{fig:WASP189_pp_hi}. Using our synthetic observations, we report detections for Fe, Fe$^+$  Ti, Ti$^+$ and V that are consistent with the detections in \citet{Prinoth2022}. \citet{Prinoth2022} also report detections of TiO, which is not seen in the synthetic model detections of WASP-189b.

\section{Discussion}
\label{sec:discussion}

As discussed in \citet{Bell2018}, the thermal disassociation and recombination of H$_{2}$ and H is an important factor when considering the dynamics of UHJs, an effect we do not consider in the current study.
In \citet{Tan2019}, this effect was included in GCM models and was found to have a significant impact on the thermal and dynamical structure of the atmosphere.
Future studies using GCMs in this regime may wish to adopt the \citet{Tan2019} methodology.

We have neglected the impact of shorter wavelength radiation than 0.2 $\mu$m which may also affect the thermal structure of the atmosphere, especially at very low pressure regions.
\citet{Lothringer2020} and \citet{Lothringer2020b} have shown that the very low pressure regions of the atmosphere can be affected greatly by the presence of FUV absorbing metals and molecules.
This was also seen in \citet{Lee2022}, where brown dwarf atmospheres orbiting white dwarf stars were simulated, but a lack of FUV opacities in the correlated-k scheme failed to reproduce well the very low pressure T-p structures of the 1D RCE models.
Important sources of opacity in this wavelength regime include H$_{2}$, SiO, H$_{2}$O, CO, H$_{2}$S, H$^{-}$ and atomic bound free opacity \citep{Sharp2007}.
Future studies may wish to include a few FUV bands to capture this effect, we suggest a practical wavelength stoppage point would be 0.13 $\mu$m to better capture the absorption of UV by CO, before H$_{2}$ dominates the UV opacity.
This is likely to be sufficient for GCM modelling purposes, since significant H$_{2}$ absorption is likely to occur at pressures below those considered here (10$^{-6}$ bar), as well as H$_{2}$ being most probably thermally and/or photochemically dissociated in these regions \citep[e.g.][]{Tsai2021}.
Despite this, our models are able to reproduce well the low pressure steep temperature inversions occurring near 10$^{-4}$ bar as seen in the 1D models of \citet{Lothringer2020b}.

For WASP-121b, our post-processing of the GCM model from \citet{Parmentier2018} provides a better fit to the low resolution and photometry data, for all the transmission, emission and TESS phase curve.
From the SPARC/MITgcm spectra compared to the Exo-FMS, it is clear that hotter T-p profiles found in SPARC/MITgcm increase in dayside planetary flux to better fit the HST WFC3 data.
The SPARC/MITgcm model also appears to fit the transmission spectra trends found in \citet{Evans2016,Evans2018} and \citet{Wilson2021} better.
From comparing the T-p profiles between the models, the SPARC/MITgcm model is cool enough to retrain larger TiO and VO abundances at the terminators compared to the Exo-FMS model, which is too hot and dissociates TiO and VO more.
This is probably responsible for increasing the R$_{\rm p}$/R$_{\star}$ in the optical to the observed levels.

From both the WASP-121b and WASP-189b models it is clear that a significant absorption features occurs at NUV wavelengths less than 0.3$\mu$m, this is primarily attributed to gas phase SiO being present in the upper atmosphere providing a significant source of NUV absorption.
We therefore suggest that that both WASP-121b and WASP-189b are potentially excellent targets for the HST WFC3/UVIS G280 mode as demonstrated for HAT-P-41b in \citet{Wakeford2020} and for WASP-179b in \citet{Lothringer2022}.

The day-night temperature contrast is also larger in the \citet{Parmentier2018} models and also fitting better the TESS data offset from \citet{Daylan2021} compared to the Exo-FMS model.
This suggests that the simulated atmospheric drag is stronger in the SPARC/MITgcm model compared to the Exo-FMS model.
This is possibly due to the use of a Shapiro filter in SPARC/MITgcm \citep[e.g.][]{Koll2018}, which acts as a proxy hyperdiffusion term adding drag to the numerical scheme and therefore lowering the jet speeds as well as increasing its latitudinal extent \citep[e.g.][]{Heng2011}.
The current Exo-FMS HJ models only use divergence dampening to stabilise the simulation which does not affect the equatorial jet, as it has zero divergence, meaning the SPARC/MITgcm model has inherently more numerical drag included in it's flow stabilisation scheme compared to the Exo-FMS (\citetalias{Hammond2022} \citeyear{Hammond2022}).

Such high-resolution spectral modelling may aid in the selection of `correct' numerical dissipation, stabilisation and drag schemes used in HJ GCM modelling.
By experimenting and tuning the wind speeds of models using drag and stabilisation schemes, it may be possible to more accurately prescribe drag settings for particular objects.
The precise prescription of drag properties in GCM modelling is discussed in detail in \citet{Heng2011} and \citetalias{Hammond2022} \citeyear{Hammond2022}.
However, \citet{Beltz2021} show that the wind speed and rotation rate play a secondary role to the 3D temperature structure when making detection predictions using high-resolution emission spectra.
Despite this, this avenue of investigation is worth further research.

Due to these differences, we therefore propose that upper atmospheric drag, possibly from magnetic sources, may play vital role in shaping the atmospheric dynamics of UHJs where the electron abundances can be significant, even from thermal dissociation of metals alone \citep[e.g.][]{Parmentier2018} as well as photo-chemical effects \citep[e.g.][]{Lavvas2014, Helling2021}.
Magnetic drag effects have been investigated as possible source of atmospheric drag since \citet{Perna2010}.
Recently, \citet{Beltz2022} investigated the role of magnetic drag across a wider planetary parameter regime and found that even small magnetic fields strengths can alter the dynamical properties of the atmosphere and day-night contrasts significantly.
Overall, our comparison here shows that different GCMs can expect varying results depending on the numerical scheme used for stability of the simulation.
We also note the integration times of the \citet{Parmentier2018} and \citet{Mikal-Evans2022} GCM models of $\approx$ 300 and 80 days respectively.
Comparison between \citet{Parmentier2018} and \citet{Mikal-Evans2022} and our GCM models may therefore be not a fair one to draw conclusive reasons as to why different results are produced.

The cooler T-p profiles found in \citet{Parmentier2018} also allow for the condensation of potentially CaTiO$_{3}$, Al$_{2}$O$_{3}$ and Fe clouds on the nightside, while in the Exo-FMS models, due to the higher energy transport efficiency results in too hot a nightside to condense these species.
This may allow a more definitive of a truly cloud free planet where no T-p profiles cross any condensation boundaries, even on the nightside.
This clearing of nightside clouds may explain the uptick in Spitzer nightside brightness temperatures beyond T$_{\rm eq}$ $\sim$ 2250 K found in \citet{Beatty2019}, with the specifics of the nightside cloud structures modelled more extensively in \citet{Gao2021}.
Due to our post-processing results of both planets, we propose that weak TiO and VO signals in transit spectra (at low or high-resolution) from the thermal dissociation of these molecules may be a strong indication of a truly or relatively cloud free atmosphere.

Our modelled T-p structures show a dynamically distinct region at very low pressures ($\approx$10$^{-4}$ bar) forming due to the UV forcing.
This region also exhibits more isothermal temperature behaviour, suggesting that the assumption of isotherms used in high-resolution studies may be a justifiable approach.
However, this also suggests that observational methods that probe these pressures such as optical wavelength high-resolution instruments may be inferring properties of a detached layer, distinct from the dynamical and temperature regimes at lower altitudes.
Therefore, physical interpretation of such high-resolution data should keep this possibility in mind.

For WASP-121b, our models produce predictions of Fe and TiO detections.
The HARPS transmission spectra analysed by \citet{Hoeijmakers2020} resulted in detections of Fe and V (but not Ti or TiO), with our models predicting that V is not visible in the dayside emission spectrum, which is currently being investigated by Hoeijmakers et al. (in prep).
However, the TiO line list (ToTo) from ExoMol \citep{McKemmish2019} was updated to be more accurate with respect to line-position only recently (L. McKemmish per. com. August 2021) which may change future confidence in the detection of TiO with the cross-correlation technique.
From the T-p results of both the SPARC/MITgcm and Exo-FMS GCM models, the Ti bearing species are unlikely to be cold trapped in the deep atmosphere (Fig. \ref{fig:WASP121_GCM_comp}).
However, cloud modelling performed in \citet{Helling2021}, who post-processed the \citet{Parmentier2018} GCM model, suggest that the nightside hemisphere temperatures may allow small sub-micron cloud particles to form in the upper atmosphere.
Coupled cloud formation models to the GCM in the UHJ regime \citep[e.g.][]{Komacek2022} will be enlightening to see if TiO can be cold trapped in the case of WASP-121b.
Potential cold trapping of TiO was investigated for HD 209458b in \citet{Parmentier2013}.

Another explanation may be the GCM models are producing too cool an upper atmosphere for WASP-121b, which does not allow TiO (and VO) to dissociate into Ti (and V) in a large enough mixing ratio in the upper atmosphere.
Otherwise, a more simpler explanation is that enhanced metallicity from Solar for WASP-121b could boost the atomic signals.

Overall, our analysis suggests that, for WASP-121b at least, low-resolution trends may be fit adequately by the GCM models but high-resolution modelling may produce different results to the observations.
This information extracts more physical information from the GCM output than just low-resolution spectra comparisons alone, allowing modellers to see how well fit the wind speeds and temperatures are at low pressure regions in their simulations.
Modellers can then be more informed and adjust modelling techniques to understand physical processes that shape the whole 3D atmosphere of UHJs.

For WASP-189b, \citet{Prinoth2022} detected Fe, Fe$^+$, Ti, Ti$^+$, Cr, Mg, V, Mn and TiO, with tentative detections of Cr$^+$, Sc$^+$, Na, Ni and Ca.
Whereas in our modelling effort only Fe, Fe$^{+}$  Ti, Ti$^{+}$ and V were able to be conclusively detected, showing TiO as the only non-detected species in the model compared to the detection in the observations.
This suggests that the temperature in the GCM may be too hot at very low-pressure, dissociating TiO too much into atomic Ti.
However, overall, the offsets in vsys for each species detection fit well the observational data, suggesting that the GCM atmospheric winds are in the correct magnitude.

\section{Summary and Conclusions}
\label{sec:conclusions}

We have studied the atmospheric properties of the UHJs WASP-121b and WASP-189b using 3D GCM models coupled with a high-temperature correlated-k radiative-transfer scheme.
We added for the first time important UV and optical species opacities to the RT schemes of UHJs, namely TiO, SiO, Fe and Fe$^{+}$, showing that these opacities and the UV forcing have a large impact on the low pressure atmospheric properties of these planets, producing steep high altitude thermal inversions and strong zonal upper atmosphere winds.
Due to this UV absorption, our models produce a distinct very low pressure dynamical wave driven regime, unexpected from atmospheric theory.
This dynamical layer is most likely probed by the strong optical lines at high resolution, suggesting that optical wavelength cross-correlation studies are sensitive to this detached region rather than the deeper jet forming layers.

Overall, our models are able to reproduce well the T-p structure features found in 1D RCE models that include UV opacities \citep{Lothringer2020, Lothringer2020b}, suggesting that adding Fe, Fe$^{+}$, SiO, TiO and VO to our opacity tables resulted in adequately capturing the very low pressure absorption from UV irradiation.
Considering strong UV and optical absorbers leads to very different dynamical structure compared to those without, suggesting that comprehensive investigations of UHJ atmospheres should include such species in the future.

Our post-processing at low-resolution has shown that the WASP-121b GCM from \citet{Parmentier2018} produces a better fit to observational data than our presented Exo-FMS models, despite not including the UV opacity sources in the GCM model itself.
We suggest from this model difference, that upper atmosphere magnetic drag effects or photochemical effects may be important considerations in shaping the T-p profiles and dynamical regimes for UHJs as well.
Our transmission spectra of both WASP-121b and WASP-189b suggest that they may be excellent targets for the HST WFC3/UVIS G280 mode \citep{Wakeford2020,Lothringer2022} due to their large SiO absorption features at NUV wavelengths.

At high-resolution, our GCM model post-processed for different phases are able to be used produce mock observations with synthetic cross-correlation maps and synthetic detection of molecular and atomic species.
Our results suggest that the WASP-121b dayside emission spectra will exhibit strong signals of Fe and TiO with no other species (that we tested here) detected.
This is contrast to the \citet{Hoeijmakers2020} study who did not find TiO or Ti in the transmission spectrum of WASP-121b, who suggest that TiO may have rained out in the atmosphere.
Hoeijmakers et al. (in prep) will provide the first data on the dayside composition of WASP-121b at high spectral resolution.
For WASP-189b, our results conform well with the \citet{Prinoth2022} data and analysis.
However, we were not able to synthetically detect TiO, detected in \citet{Prinoth2022} study.
This suggests that the upper atmosphere of the terminator regions in the GCM may be too warm compared to the real object and thermally dissociate TiO in the upper atmosphere to lower mixing ratios that are not detectable.

As the observational data for UHJs continues to be produced at low and high resolution, theorists and modellers can further increase our knowledge of the physical interpretation of such data using multiple axis of data comparisons.
Our current study shows that modellers can now readily add 3D high-resolution predictions and synthetic molecular/atomic detections to analyse their model output in greater depth than before.
By comparing contemporary modelling efforts to both low and high resolution observational data, modellers can more easily investigate potential missing physics or the impact of adding additional physical processes to models on the 3D high spectral resolution properties of the atmosphere.

\section*{Acknowledgements}
We thank V. Parmentier for SPARC/MITgcm output of WASP-121b.
We thank T. Daylan and A. Deline for providing TESS and CHEOPS phase curve data on WASP-121b and WASP-189b respectively.
E.K.H. Lee is supported by the SNSF Ambizione Fellowship grant (\#193448).
B.P. acknowledges financial support from the Märta and Eric Holmberg Endowment.
Plots were produced using the community open-source Python packages Matplotlib \citep{Hunter2007}, SciPy \citep{Jones2001}, and AstroPy \citep{Astropy2018}.
The HPC support staff at AOPP, University of Oxford and University of Bern are highly acknowledged.

\section*{Data availability}
All GCM data and spectra post-processed products are publicly available at zendoo: [provide zendoo  DOI and link after review].
Radiative transfer models, opacities and other products are available on the lead author's Github:  \url{https://github.com/ELeeAstro} or upon request.
gCMCRT is available on GitHub: \url{https://github.com/ELeeAstro/gCMCRT}
The cross-correlation code is available at:  \url{https://github.com/hoeijmakers/tayph/}
All other data is available upon request.




\bibliographystyle{mnras}
\DeclareRobustCommand{\VAN}[3]{#3}
\bibliography{bib2.bib} 



\clearpage

\appendix

\section{Cross-correlation templates}
\label{app:templates}
\begin{figure*}
    \centering
    \includegraphics[width=\textwidth]{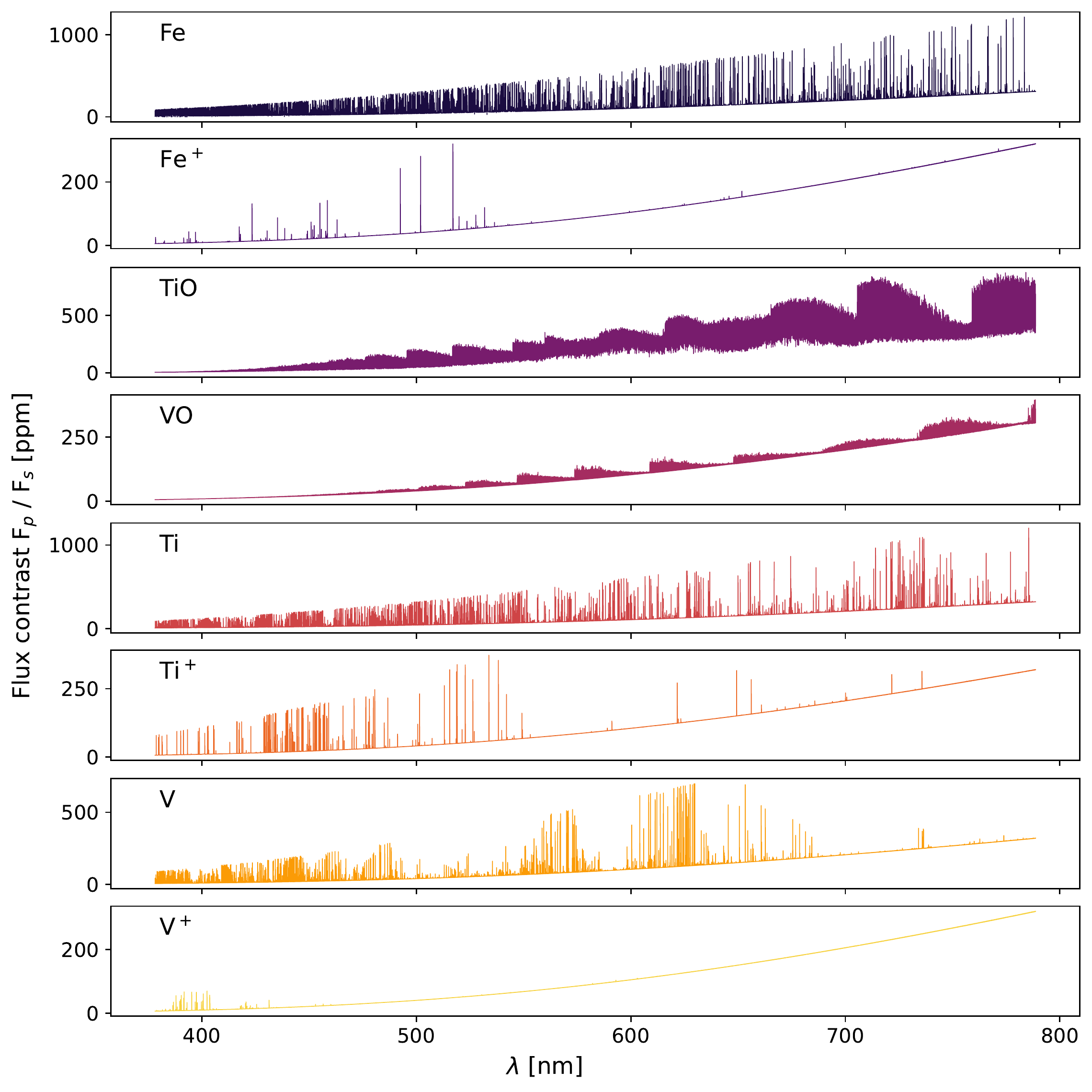}
    \caption{Cross-correlation templates in emission for WASP-121b for Fe, Fe$^{+}$, Ti, Ti$^{+}$, TiO, V, V$^{+}$ and VO. Fe$^{+}$ and V$^{+}$ are constraint to the bluer part of the ESPRESSO wavelengths, such that our detections are a result of over-predicting the SNR in the blue orders.}
    \label{fig:WASP-121b_templates}
\end{figure*}

\begin{figure*}
    \centering
    \includegraphics[width=\textwidth]{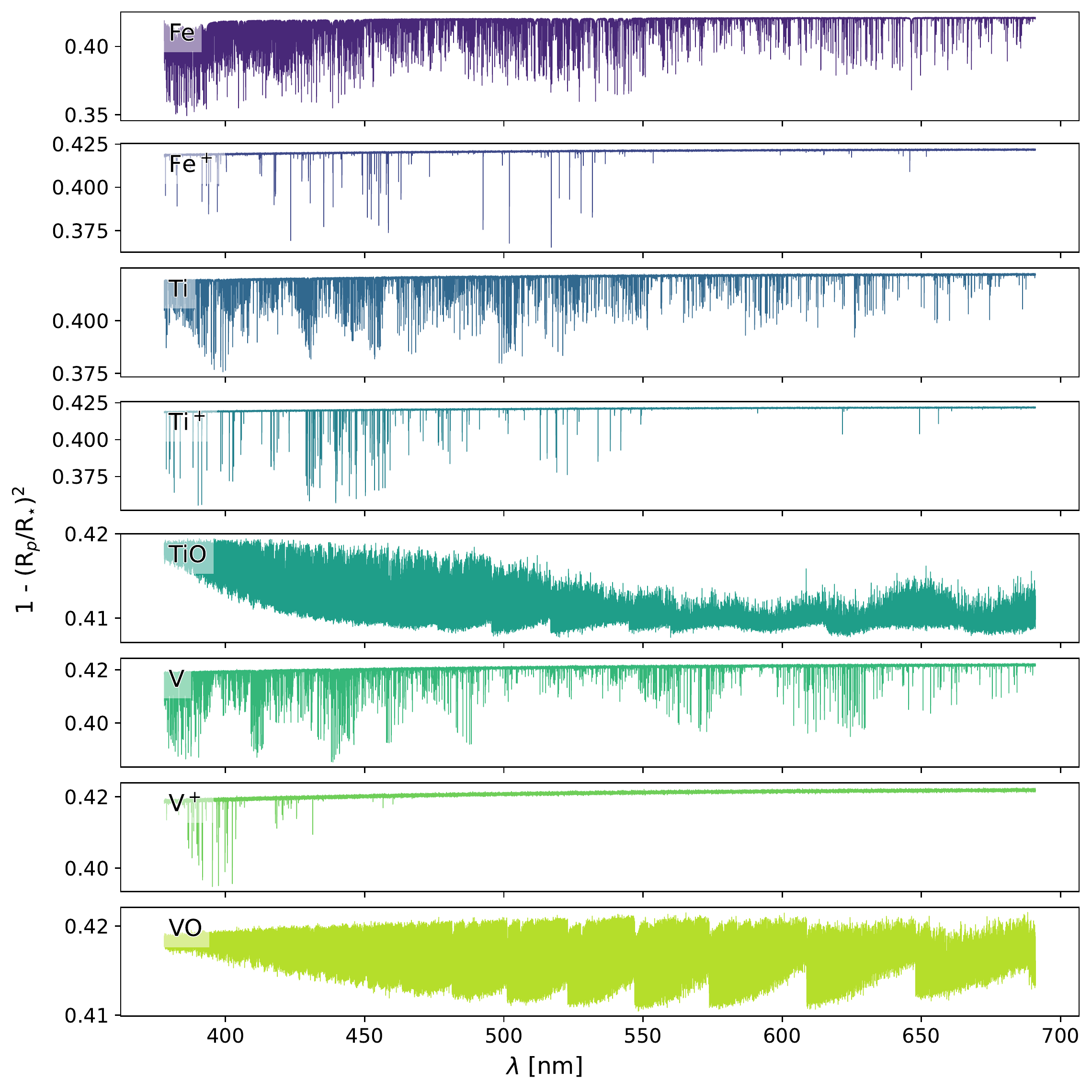}
    \caption{Cross-correlation templates in transmission for WASP-189b for Fe, Fe$^{+}$, Ti, Ti$^{+}$, TiO, V, V$^{+}$ and VO.}
    \label{fig:WASP-189b_templates}
\end{figure*}

\clearpage

\bsp	
\label{lastpage}
\end{document}